\documentclass[12pt,english,onecolumn]{IEEEtran}
\usepackage[T1]{fontenc}
\usepackage[latin1]{inputenc}
\usepackage{subfigure}
\usepackage{amsmath}
\usepackage{graphicx}
\usepackage{amssymb}

\makeatletter
 \newtheorem{thm}{Theorem}
 \newtheorem{cor}{Corollary}
 \newtheorem{example}{Example}
 \newtheorem{remrk}{Remark}
 \newtheorem{case}{Case}
 \newtheorem{lemma}{Lemma}

\usepackage{cite}
\usepackage{url}
\usepackage{bm}
\usepackage{bbm}

\usepackage{babel}
\makeatother
\begin{document}

\title{Quantization Bounds on Grassmann Manifolds and Applications to MIMO
Communications$^{^{*}}$}

\author{Wei Dai, Youjian Liu and Brian Rider\\
 University of Colorado at Boulder\\
425 UCB, Boulder, CO, 80309, USA}

\maketitle

\begin{abstract}
This paper considers the quantization problem on the Grassmann manifold
$\mathcal{G}_{n,p}$, the set of all $p$-dimensional planes (through
the origin) in the $n$-dimensional Euclidean space. The chief result
is a closed-form formula for the volume of a metric ball in the Grassmann
manifold when the radius is sufficiently small. This volume formula
holds for Grassmann manifolds with arbitrary dimension $n$ and $p$,
while previous results pertained only to $p=1$, or a fixed $p$ with
asymptotically large $n$. Based on this result, several quantization
bounds are derived for sphere packing and rate distortion tradeoff.
We establish asymptotically equivalent lower and upper bounds for
the rate distortion tradeoff. Since the upper bound is derived by
constructing random codes, this result implies that the random codes
are asymptotically optimal. The above results are also extended to
the more general case, in which $\mathcal{G}_{n,q}$ is quantized
through a code in $\mathcal{G}_{n,p}$, where $p$ and $q$ are not
necessarily the same. Finally, we discuss some applications of the
derived results to multi-antenna communication systems.
\end{abstract}
\renewcommand{\thefootnote}{\fnsymbol{footnote}} \footnotetext[1]{This work is supported by the NSF DMS-0508680, and  the Junior Faculty Development Award, University of Colorado at Boulder. It is submitted to IEEE Transactions on Information Theory, Aug. 18, 2005 and revised on April 22, 2006. Part of this work was presented at the IEEE Globe Telecommunications Conference, St. Louis, MO, Nov. 28 - Dec. 2, 2005} \renewcommand{\thefootnote}{\arabic{footnote}} \setcounter{footnote}{0}

\section{Introduction\label{sec:Introduction}}

%Introduction

The \emph{Grassmann manifold} $\mathcal{G}_{n,p}\left(\mathbb{L}\right)$
is the set of all $p$-dimensional planes (through the origin) in
the $n$-dimensional Euclidean space $\mathbb{L}^{n}$, where $\mathbb{L}$
is either $\mathbb{R}$ or $\mathbb{C}$. It forms a compact Riemann
manifold of real dimension $\beta p\left(n-p\right)$, where $\beta=1$
when $\mathbb{L}=\mathbb{R}$ and $\beta=2$ when $\mathbb{L}=\mathbb{C}$.
The Grassmann manifold provides a useful analysis tool for multi-antenna
communications (also known as multiple-input multiple-output (MIMO)
communication systems). For non-coherent MIMO systems, sphere packings
of $\mathcal{G}_{n,p}\left(\mathbb{L}\right)$ can be viewed as a
generalization of spherical codes \cite{Urbanke_IT01_Signal_Constellations,Tse_IT02_Communication_on_Grassmann_Manifold,Barg_IT02_Bounds_Grassmann_Manifold}.
For MIMO systems with partial channel state information at the transmitter
(CSIT), which is obtained by finite-rate channel-state feedback, the
quantization of beamforming matrices is related to the quantization
on the Grassmann manifold \cite{Sabharwal_IT03_Beamforming_MIMO,Love_IT03_Grassman_Beamforming_MIMO,Dai_05_Power_onoff_strategy_design_finite_rate_feedback}. 

The basic quantization problems addressed in this paper are the sphere
packing bounds and rate distortion tradeoff. Roughly speaking, a quantization
is a representation of a source in $\mathcal{G}_{n,p}\left(\mathbb{L}\right)$.
In particular, it maps an element in $\mathcal{G}_{n,p}\left(\mathbb{L}\right)$
into a subset of $\mathcal{G}_{n,p}\left(\mathbb{L}\right)$, known
as a code $\mathcal{C}$. Define the minimum distance of a code $\delta\triangleq\delta\left(\mathcal{C}\right)$
as the minimum distance between any two codewords in the code $\mathcal{C}$.
A sphere packing bound relates the size of a code and a given minimum
distance $\delta$. Rate distortion tradeoff is another important
aspect of the quantization problem. A distortion metric is a mapping
from the set of element pairs in $\mathcal{G}_{n,p}\left(\mathbb{L}\right)$
into the set of non-negative real numbers. Given a source distribution
and a distortion metric, the rate distortion tradeoff is described
by the minimum expected distortion achievable for a given code size,
or equivalently the minimum code size required to achieve a particular
expected distortion.

There are several papers addressing the quantization problem for Grassmann
manifolds. In \cite{Conway_96_PackingLinesPlanes}, an isometric embedding
of $\mathcal{G}_{n,p}\left(\mathbb{R}\right)$ into a sphere in Euclidean
space $\mathbb{R}^{\frac{1}{2}\left(m-1\right)\left(m+2\right)}$
is given. Then, using the Rankin bound in Euclidean space, the Rankin
bound in $\mathcal{G}_{n,p}\left(\mathbb{R}\right)$ is obtained.
Unfortunately, this bound is not tight when the code size is large.
Instead of resorting to an isometric embedding, sphere packing bounds
can also be derived from analysis in the Grassmann manifold directly.
Let $B\left(\delta\right)$ denote a metric ball of radius $\delta$
in $\mathcal{G}_{n,p}\left(\mathbb{L}\right)$. The sphere packing
bounds can be derived from the volume of $B\left(\delta\right)$ \cite{Barg_IT02_Bounds_Grassmann_Manifold}.
The exact volume formula for a $B\left(\delta\right)$ in $\mathcal{G}_{n,p}\left(\mathbb{L}\right)$
with $p=1$ and $\mathbb{L}=\mathbb{C}$ is derived in \cite{Sabharwal_IT03_Beamforming_MIMO}.
An asymptotic volume formula for a $B\left(\delta\right)$ in $\mathcal{G}_{n,p}\left(\mathbb{L}\right)$,
where $p\geq1$ is fixed and $n$ approaches infinity, is derived
in \cite{Barg_IT02_Bounds_Grassmann_Manifold}. Based on those volume
formulas, the corresponding sphere packing bounds are developed in
\cite{Love_IT03_Grassman_Beamforming_MIMO,Barg_IT02_Bounds_Grassmann_Manifold}.
Besides the sphere packing bounds, the rate distortion tradeoff is
also treated in \cite{Heath_ICASSP05_Quantization_Grassmann_Manifold},
where approximations to the distortion rate function are derived via
the sphere packing bounds on the Grassmann manifold. However, the
derived approximations are based on the volume formulas in \cite{Barg_IT02_Bounds_Grassmann_Manifold,Sabharwal_IT03_Beamforming_MIMO}
which are only valid for some special choices of $n$ and $p$: either
$p=1$ or fixed $p\geq1$ with asymptotic large $n$.

The main contribution of this paper is to derive a closed-form formula
for the volume of a small ball in the Grassmann manifold. Based on
this formula, sphere packing bounds are derived and rate distortion
tradeoff are accurately quantified. Specifically:

\begin{enumerate}
\item An explicit volume formula for a metric ball in $\mathcal{G}_{n,p}\left(\mathbb{L}\right)$
is derived when the radius $\delta$ is sufficiently small. It holds
for Grassmann manifolds with arbitrary dimensions while previous results
are only valid for either $p=1$ or a fixed $p$ with asymptotically
large $n$. The main order term of the volume is $c_{n,p,p,\beta}\delta^{\beta p\left(n-p\right)}$
for a constant $c_{n,p,p,\beta}$ depending on $n$, $p$ and $\beta$.
Lower and upper bounds on the volume formula are also derived. 
\item Based on the volume formula, the Gilbert-Varshamov and Hamming bounds
for sphere packings are obtained. For the distortion rate function,
a lower bound is established via sphere packing argument and an upper
bound is derived via random-code argument. The bounds are in fact
asymptotically identical, and so precisely quantify the asymptotic
rate distortion tradeoff. Since the upper bound is actually derived
from the average distortion of random codes, it follows that random
codes are asymptotically optimal.
\item The volume formula and the results on the rate distortion tradeoff
are extended to a more general plane matching problem. In this plane
matching problem, a plane from the code $\mathcal{C}\subset\mathcal{G}_{n,p}\left(\mathbb{L}\right)$
is chosen to match a random plane $Q\in\mathcal{G}_{n,q}\left(\mathbb{L}\right)$
to minimize the distortion, where $p$ and $q$ are not necessarily
the same. For plane matching, a metric ball in $\mathcal{G}_{n,q}\left(\mathbb{L}\right)$
centered at a plane in $\mathcal{G}_{n,p}\left(\mathbb{L}\right)$
is studied. The volume formula is derived for such a ball with sufficiently
small radius. The rate distortion tradeoff is also quantified by the
same method as above.
\item As an application of the derived quantization bounds, the information
rate of a MIMO system with finite-rate channel-state feedback and
power on/off strategy is accurately quantified for the first time.
Since the corresponding Grassmann manifold for most practical MIMO
systems has $p>1$ and small $n$, the quantization bounds derived
in this paper are necessary.
\end{enumerate}

The paper is organized as follows. Section \ref{sec:Preliminaries}
provides some preliminaries on the Grassmann manifold. Section \ref{sec:Spheres}
derives the explicit volume formula for a metric ball in the Grassmann
manifold. The corresponding sphere packing bounds are obtained and
the rate distortion tradeoff is accurately quantified in Section \ref{sec:Quantization-Bounds}.
An application of the quantization bounds to MIMO systems with finite-rate
channel-state feedback is detailed in Section \ref{sec:Application}.
Section \ref{sec:Conclusion} contains the conclusions.

\section{\label{sec:Preliminaries}Preliminaries}

This section presents a brief introduction to the Grassmann manifold.
A metric and a measure on the Grassmann manifold are defined, and
the problems relevant to quantization on the Grassmann manifold are
formulated. For completeness, we also extend the quantization problem
to a more general plane matching problem.

\subsection{\label{sub:Metric-and-Measure}Metric and Measure on $\mathcal{G}_{n,p}\left(\mathbb{L}\right)$}

For the sake of applications \cite{Sabharwal_IT03_Beamforming_MIMO,Love_IT03_Grassman_Beamforming_MIMO,Dai_05_Power_onoff_strategy_design_finite_rate_feedback},
the projection Frobenius metric (\emph{chordal distance}) is employed
throughout the paper although the corresponding analysis is also applicable
to the geodesic metric \cite{Barg_IT02_Bounds_Grassmann_Manifold}.
For any two planes $P,Q\in\mathcal{G}_{n,p}\left(\mathbb{L}\right)$,
we define the principle angles and the chordal distance between $P$
and $Q$ as follows. Let $\mathbf{u}_{1}\in P$ and $\mathbf{v}_{1}\in Q$
be the unit vectors such that $\left|\mathbf{u}_{1}^{\dagger}\mathbf{v}_{1}\right|$
is maximal. Inductively, let $\mathbf{u}_{i}\in P$ and $\mathbf{v}_{i}\in Q$
be the unit vectors such that $\mathbf{u}_{i}^{\dagger}\mathbf{u}_{j}=0$
and $\mathbf{v}_{i}^{\dagger}\mathbf{v}_{j}=0$ for all $1\leq j<i$
and $\left|\mathbf{u}_{i}^{\dagger}\mathbf{v}_{i}\right|$ is maximal.
The principle angles are defined as $\theta_{i}=\arccos\left|\mathbf{u}_{i}^{\dagger}\mathbf{v}_{i}\right|$
for $i=1,\cdots,p$ \cite{Conway_96_PackingLinesPlanes,James_54_Normal_Multivariate_Analysis_Orthogonal_Group}.
The chordal distance between $P$ and $Q$ is given by \begin{equation}
d_{c}\left(P,Q\right)\triangleq\sqrt{\sum_{i=1}^{p}\sin^{2}\theta_{i}}.\label{eq:def-dc}\end{equation}

The invariant measure $\mu$ on $\mathcal{G}_{n,p}\left(\mathbb{L}\right)$
is defined as follows. Let $O\left(n\right)$ and $U\left(n\right)$
be the groups of $n\times n$ orthogonal and unitary matrices respectively.
Let $\mathbf{A},\mathbf{B}\in O\left(n\right)$ when $\mathbb{L}=\mathbb{R}$,
or $\mathbf{A},\mathbf{B}\in U\left(n\right)$ when $\mathbb{L}=\mathbb{C}$.
For any measurable set $\mathcal{M}\subset\mathcal{G}_{n,p}\left(\mathbb{L}\right)$
and arbitrary $\mathbf{A}$ and $\mathbf{B}$, \[
\mu\left(\mathbf{A}\mathcal{M}\right)=\mu\left(\mathcal{M}\right)=\mu\left(\mathcal{M}\mathbf{B}\right).\]
The invariant measure defines the uniform/isotropic distribution on
$\mathcal{G}_{n,p}\left(\mathbb{L}\right)$ as well \cite{James_54_Normal_Multivariate_Analysis_Orthogonal_Group}.

\subsection{\label{sub:Quantization-Problem}Quantization on $\mathcal{G}_{n,p}\left(\mathbb{L}\right)$}

Given both a metric and a measure on $\mathcal{G}_{n,p}\left(\mathbb{L}\right)$,
a quantization on the Grassmann manifold can be well defined. Let
$\mathcal{C}$ be a finite size discrete subset of $\mathcal{G}_{n,p}\left(\mathbb{L}\right)$.
A quantization is a mapping from the $\mathcal{G}_{n,p}\left(\mathbb{L}\right)$
to the set $\mathcal{C}$ (also known as a code), i.e., \[
\mathfrak{q}:\mathcal{G}_{n,p}\left(\mathbb{L}\right)\rightarrow\mathcal{C}.\]
An element in the code $\mathcal{C}$ is called a codeword. Thus,
roughly speaking, a quantization is to use a subset of $\mathcal{G}_{n,p}\left(\mathbb{L}\right)$
to represent the whole space.

Sphere packing bounds relate the size of the code to the minimum distance
among the codewords. Let $\delta$ be the minimum distance between
any two codewords of a code $\mathcal{C}$ and $B\left(\delta\right)$
be a metric ball of radius $\delta$ in the $\mathcal{G}_{n,p}\left(\mathbb{L}\right)$.
If $K$ is any positive integer such that $K\mu\left(B\left(\delta\right)\right)<1$,
then there exists a code $\mathcal{C}$ of size $K+1$ with minimum
distance $\delta$. This principle is called as the \emph{Gilbert-Varshamov}
lower bound, \begin{equation}
\left|\mathcal{C}\right|>\frac{1}{\mu\left(B\left(\delta\right)\right)}.\label{eq:GV_bd}\end{equation}
On the other hand, $\left|\mathcal{C}\right|\mu\left(B\left(\delta/2\right)\right)\leq1$
for any code $\mathcal{C}$. The \emph{Hamming} upper bound captures
this fact as \begin{equation}
\left|\mathcal{C}\right|\leq\frac{1}{\mu\left(B\left(\delta/2\right)\right)}.\label{eq:Hamming_bd}\end{equation}
For more information about the sphere packing bounds, see \cite{Barg_IT02_Bounds_Grassmann_Manifold}. 

Rate distortion tradeoff is another important aspect of the quantization
problem. A distortion metric is a mapping, \[
\mathfrak{d}:\mathcal{G}_{n,p}\left(\mathbb{L}\right)\times\mathcal{C}\rightarrow\left[0,+\infty\right),\]
from the set of the element pairs in $\mathcal{G}_{n,p}\left(\mathbb{L}\right)$
and $\mathcal{C}$ into the set of non-negative real numbers. Throughout
this paper, we define the distortion metric as the square of the chordal
distance, $\mathfrak{d}\left(\cdot,\cdot\right)\triangleq d_{c}^{2}\left(\cdot,\cdot\right)$.
Assume that a source $Q$ is randomly distributed in $\mathcal{G}_{n,p}\left(\mathbb{L}\right)$.
The distortion associated with a quantization $\mathfrak{q}$ is defined
as \[
D\triangleq\mathrm{E}\left[\mathfrak{d}\left(Q,\mathfrak{q}\left(Q\right)\right)\right]=\mathrm{E}\left[d_{c}^{2}\left(Q,\mathfrak{q}\left(Q\right)\right)\right].\]
The \emph{rate distortion tradeoff} can be described by the infimum
achievable distortion given a code size, which is called the \emph{distortion
rate function}, or equivalently the infimum code size required to
achieve a particular distortion, which is called the \emph{rate distortion
function}. In this paper, the source $Q$ is assumed to be uniformly
distributed in $\mathcal{G}_{n,p}\left(\mathbb{L}\right)$. For a
given code $\mathcal{C}\subset\mathcal{G}_{n,p}\left(\mathbb{L}\right)$,
the optimal quantization to minimize the distortion is given by%
\footnote{The ties, i.e. the case that $\exists P_{1},P_{2}\in\mathcal{C}$
such that $d_{c}\left(P_{1},Q\right)=\underset{P\in\mathcal{C}}{\min}\; d_{c}\left(P,Q\right)=d_{c}\left(P_{2},Q\right)$,
are broken arbitrarily as they occur with probability zero.%
} \[
\mathfrak{q}\left(Q\right)=\arg\;\underset{P\in\mathcal{C}}{\min}\; d_{c}\left(P,Q\right).\]
The distortion associated with this quantization is \begin{eqnarray*}
D\left(\mathcal{C}\right) & = & \mathrm{E}\left[\underset{P\in\mathcal{C}}{\min}\; d_{c}^{2}\left(P,Q\right)\right].\end{eqnarray*}
For a given code size $K$ where $K$ is a positive integer, the distortion
rate function is then given by%
\footnote{The standard definition of the distortion rate function involves the
code rate, which is $\log_{2}K$. The definition in this paper is
equivalent to the standard one.%
} \begin{equation}
D^{*}\left(K\right)=\underset{\mathcal{C}:\left|\mathcal{C}\right|=K}{\inf}\; D\left(\mathcal{C}\right).\label{eq:distor-rate-fn}\end{equation}
 The rate distortion function is given by \begin{equation}
K^{*}\left(D\right)=\underset{D\left(\mathcal{C}\right)\leq D}{\inf}\;\left|\mathcal{C}\right|.\label{eq:rate-distor-fn}\end{equation}

\subsection{\label{sub:Plane-Matching-Problem}An Extension: Plane Matching Problem}

For the sake of completeness, we extend the quantization problem to
a more general plane matching problem. The plane matching problem
involves planes from different spaces $\mathcal{G}_{n,p}\left(\mathbb{L}\right)$
and $\mathcal{G}_{n,q}\left(\mathbb{L}\right)$ where $p$ and $q$
are not necessarily the same. 

To formulate the plane matching problem, we need to define the chordal
distance between $P\in\mathcal{G}_{n,p}\left(\mathbb{L}\right)$ and
$Q\in\mathcal{G}_{n,q}\left(\mathbb{L}\right)$. Without loss of generality,
we assume that $p\leq q$. Using the same procedure described in Section
\ref{sub:Metric-and-Measure}, we are able to define the principle
angles $0\leq\theta_{1}\leq\cdots\leq\theta_{p}\leq\frac{\pi}{2}$.
Based on the principle angles, the chordal distance between $P\in\mathcal{G}_{n,p}\left(\mathbb{L}\right)$
and $Q\in\mathcal{G}_{n,q}\left(\mathbb{L}\right)$ are defined as
$d_{c}\left(P,Q\right)\triangleq\sqrt{\sum_{i=1}^{p}\sin^{2}\theta_{i}}$.
In this way, the definition of chordal distance in (\ref{eq:def-dc})
is just a particular case of the general definition.

Now consider the plane matching problem. Intuitively, the plane matching
problem is to choose a plane from the code $\mathcal{C}\subset\mathcal{G}_{n,p}\left(\mathbb{L}\right)$
to match a random plane $Q\in\mathcal{G}_{n,q}\left(\mathbb{L}\right)$
such that the average distortion is minimized, where $1\leq p\leq n$
and $1\le q\le n$ are not necessarily the same. Formally, a plane
matching is a map from the whole space of Grassmann manifold, e.g.,
$\mathcal{G}_{n,q}\left(\mathbb{L}\right)$, to the code $\mathcal{C}\subset\mathcal{G}_{n,p}\left(\mathbb{L}\right)$,
\[
\mathfrak{q}:\mathcal{G}_{n,q}\left(\mathbb{L}\right)\rightarrow\mathcal{C},\]
such that \[
D\left(\mathcal{C}\right)\triangleq\mathrm{E}_{Q}\left[\underset{P\in\mathcal{C}}{\min}\; d_{c}^{2}\left(P,Q\right)\right]\]
is minimized. According to the same principles in the quantization
problem, the rate distortion tradeoff can be extended to the plane
matching problem.

\section{\label{sec:Spheres}Metric Balls in the Grassmann Manifold}

In this section, an explicit volume formula for a metric ball $B\left(\delta\right)$
in the Grassmann manifold is derived. It is the essential tool to
quantify the rate distortion tradeoff in Section \ref{sec:Quantization-Bounds}.

The volume calculation depends on the relationship between the measure
and the metric defined on the Grassmann manifold. This paper focuses
on the invariant measure $\mu$, which corresponds to the uniform/isotropic
distribution, and the chordal distance $d_{c}$. For any given $P\in\mathcal{G}_{n,p}\left(\mathbb{L}\right)$
and $Q\in\mathcal{G}_{n,q}\left(\mathbb{L}\right)$, define \[
B_{P}\left(\delta\right)=\left\{ \hat{Q}\in\mathcal{G}_{n,q}\left(\mathbb{L}\right):\; d_{c}\left(P,\hat{Q}\right)\leq\delta\right\} \]
and \[
B_{Q}\left(\delta\right)=\left\{ \hat{P}\in\mathcal{G}_{n,p}\left(\mathbb{L}\right):\; d_{c}\left(\hat{P},Q\right)\leq\delta\right\} .\]
For the invariant measure $\mu$, it has been shown that $\mu\left(B_{P}\left(\delta\right)\right)=\mu\left(B_{Q}\left(\delta\right)\right)$
and the value is independent of the choice of the center \cite{James_54_Normal_Multivariate_Analysis_Orthogonal_Group}.
It is convenient to denote $B_{P}\left(\delta\right)$ and $B_{Q}\left(\delta\right)$
by $B\left(\delta\right)$ without distinguishing them. Then, the
volume of a metric ball $B\left(\delta\right)$ is given by\begin{equation}
\mu\left(B\left(\delta\right)\right)=\underset{\sum_{i=1}^{p}\sin^{2}\theta_{i}\leq\delta^{2}}{\int\cdots\int}\; d\mu_{\bm{\theta}},\label{eq:actual_volume}\end{equation}
where $1\leq\theta_{1}\leq\frac{\pi}{2},\cdots,1\leq\theta_{p}\leq\frac{\pi}{2}$
are the principle angles and the differential form $d\mu_{\bm{\theta}}$
is the joint density of the $\theta_{i}$'s, which is given in \cite{James_54_Normal_Multivariate_Analysis_Orthogonal_Group,Muirhead_book82_multivariate_statistics,Adler_2001_Integrals_Grassmann}
and as well (\ref{eq:d_mu_0}) in Appendix \ref{sub:Proof-of-Volume-Formula}
below. 

The following theorem calculates the volume formula and expresses
it as an exponentiation of the radius.

\begin{thm}
\label{thm:Volume_formula}When $\delta\leq1$, the volume of a metric
ball $B\left(\delta\right)$ is given by \begin{equation}
\mu\left(B\left(\delta\right)\right)=c_{n,p,q,\beta}\delta^{\beta p\left(n-q\right)}\left(1+c_{n,p,q,\beta}^{\left(1\right)}\delta^{2}+o\left(\delta^{2}\right)\right),\label{eq:simplified_volume_formula}\end{equation}
where\[
\beta=\left\{ \begin{array}{ll}
1 & \mathrm{if}\;\mathbb{L}=\mathbb{R}\\
2 & \mathrm{if}\;\mathbb{L}=\mathbb{C}\end{array}\right.,\]
\begin{equation}
c_{n,p,q,\beta}=\left\{ \begin{array}{ll}
\frac{1}{\Gamma\left(\frac{\beta}{2}p\left(n-q\right)+1\right)}\prod_{i=1}^{p}\frac{\Gamma\left(\frac{\beta}{2}\left(n-i+1\right)\right)}{\Gamma\left(\frac{\beta}{2}\left(q-i+1\right)\right)} & \mathrm{if}\;1\leq p\leq q\leq\frac{n}{2}\\
\frac{1}{\Gamma\left(\frac{\beta}{2}p\left(n-q\right)+1\right)}\prod_{i=1}^{p}\frac{\Gamma\left(\frac{\beta}{2}\left(n-i+1\right)\right)}{\Gamma\left(\frac{\beta}{2}\left(n-p-i+1\right)\right)} & \mathrm{if}\;1\leq p\leq\frac{n}{2}\leq q\leq n\;\mathrm{and}\; p+q\leq n\\
\frac{1}{\Gamma\left(\frac{\beta}{2}p\left(n-q\right)+1\right)}\prod_{i=1}^{n-q}\frac{\Gamma\left(\frac{\beta}{2}\left(n-i+1\right)\right)}{\Gamma\left(\frac{\beta}{2}\left(q-i+1\right)\right)} & \mathrm{if}\;1\leq p\leq\frac{n}{2}\leq q\leq n\;\mathrm{and}\; p+q\geq n\\
\frac{1}{\Gamma\left(\frac{\beta}{2}p\left(n-q\right)+1\right)}\prod_{i=1}^{n-q}\frac{\Gamma\left(\frac{\beta}{2}\left(n-i+1\right)\right)}{\Gamma\left(\frac{\beta}{2}\left(n-p-i+1\right)\right)} & \mathrm{if}\;\frac{n}{2}\leq p\leq q\leq n\end{array}\right.,\label{eq:constant-0}\end{equation}
and \begin{equation}
c_{n,p,q,\beta}^{\left(1\right)}=-\left(\frac{\beta}{2}\left(q-p+1\right)-1\right)\frac{\frac{\beta}{2}p\left(n-q\right)}{\frac{\beta}{2}p\left(n-q\right)+1}.\label{eq:constant-1}\end{equation}

\end{thm}
\begin{proof}
See Appendix \ref{sub:Proof-of-Volume-Formula}.
\end{proof}
%Theorem

The following corollary gives the two cases where the volume formula
becomes exact.

\begin{cor}
\label{cor:Exact-Volume-Formula}When $\delta\leq1$, in either of
the following two cases, 
\begin{enumerate}
\item $\mathbb{L}=\mathbb{C}$ and $q=p$;
\item $\mathbb{L}=\mathbb{R}$ and $q=p+1$, 
\end{enumerate}
the volume of a metric ball $B\left(\delta\right)$ can be exactly
calculated by\[
\mu\left(B\left(\delta\right)\right)=c_{n,p,q,\beta}\delta^{\beta p\left(n-q\right)},\]
 where $c_{n,p,q,\beta}$ is defined in (\ref{eq:constant-0}). 
\end{cor}

We also have the general bounds:

\begin{cor}
\label{cor:volume-bounds}Assume $\delta\leq1$. If $\mathbb{L}=\mathbb{R}$
and $p=q$ , the volume of $B\left(\delta\right)$ is bounded by \[
c_{n,p,p,1}\delta^{p\left(n-p\right)}\leq\mu\left(B\left(\delta\right)\right)\leq c_{n,p,p,1}\delta^{p\left(n-p\right)}\left(1-\delta^{2}\right)^{-\frac{p}{2}}.\]
For all other cases, \[
\left(1-\delta^{2}\right)^{\frac{\beta}{2}p\left(q-p+1\right)-p}c_{n,p,q,\beta}\delta^{\beta p\left(n-q\right)}\leq\mu\left(B\left(\delta\right)\right)\leq c_{n,p,q,\beta}\delta^{\beta p\left(n-q\right)}.\]

\end{cor}
\begin{proof}
Corollary \ref{cor:Exact-Volume-Formula} and \ref{cor:volume-bounds}
follow the proof of Theorem \ref{thm:Volume_formula} by tracking
the higher order terms. 
\end{proof}

Theorem \ref{thm:Volume_formula} is of course consistent with the
previous results in \cite{Sabharwal_IT03_Beamforming_MIMO} and \cite{Barg_IT02_Bounds_Grassmann_Manifold},
which pertain to special choices of $n$ and $p$ and are stated below
as examples.

\begin{example}
\label{ex:G_n_1}Consider the volume formula for a $B\left(\delta\right)$
where $p=q=1$. Without normalization, the total volume of $\mathcal{G}_{n,1}\left(\mathbb{C}\right)$
is $2\pi^{n}/\left(n-1\right)!$ and the volume of the $B\left(\delta\right)$
is $2\pi^{n}\delta^{2\left(n-1\right)}/\left(n-1\right)!$ \cite{Sabharwal_IT03_Beamforming_MIMO}.
Therefore, \[
\mu\left(B\left(\delta\right)\right)=\delta^{2\left(n-1\right)},\]
agreeing with Theorem \ref{thm:Volume_formula} where $\beta=2$ and
$c_{n,1,1,2}=1$.
\end{example}
%Gn,1

\begin{example}
\label{ex:G_asymptotic_n}For the case that $p=q$ are fixed and $n\rightarrow+\infty$,
an asymptotic volume formula for a $B\left(\delta\right)$ is derived
by Barg \cite{Barg_IT02_Bounds_Grassmann_Manifold}, which reads\begin{equation}
\mu\left(B\left(\delta\right)\right)=\left(\frac{\delta}{\sqrt{p}}\right)^{\beta np+o\left(n\right)}.\label{eq:Barg-formula}\end{equation}
On the other hand, asymptotic analysis from Theorem \ref{thm:Volume_formula}
gives \[
\mu\left(B\left(\delta\right)\right)=\left(\frac{\delta}{\sqrt{p}}\right)^{\beta p\left(n-q\right)+O\left(\log n\right)}\left(1-\left(\frac{\beta}{2}\left(q-p+1\right)-1\right)\delta^{2}+o\left(\delta^{2}\right)\right)\]
for asymptotically large $n$ and any fixed $1\le p\leq q$. The derivation
follows the Stirling's approximation applied to $-\frac{1}{\beta p\left(n-q\right)}\log\left(c_{n,p,q,\beta}\right)$.
In this setting, Theorem \ref{thm:Volume_formula} is consistent with
(\ref{eq:Barg-formula}) and provides refinement.
\end{example}

Importantly though, Theorem \ref{thm:Volume_formula} is distinct
from the above results in that it holds for arbitrary $p$, $q$ and
$n$. For a metric ball with parameter $n$ not asymptotically large,
i.e., $p$ and $q$ are comparable to $n$, it is not appropriate
to use (\ref{eq:Barg-formula}) to estimate the volume. A trivial
example is that the $p=q=n$ case. If $p=q=n$, the exact volume of
$B\left(\delta\right)$ for $\forall\delta>0$ is the constant $1$.
The formula in Theorem \ref{thm:Volume_formula} gives $c_{n,n,n,\beta}=1$
and $c_{n,n,n,\beta}\delta^{\beta p\left(n-q\right)}=1$. However,
the approximation $\left(\delta/\sqrt{n}\right)^{\beta n^{2}}$ (formula
(\ref{eq:Barg-formula})) will give a small number much less than
$1$ when $\delta$ is small.

\begin{figure}
\subfigure[Real Grassmann manifolds]{\includegraphics[%
  clip,
  scale=0.8]{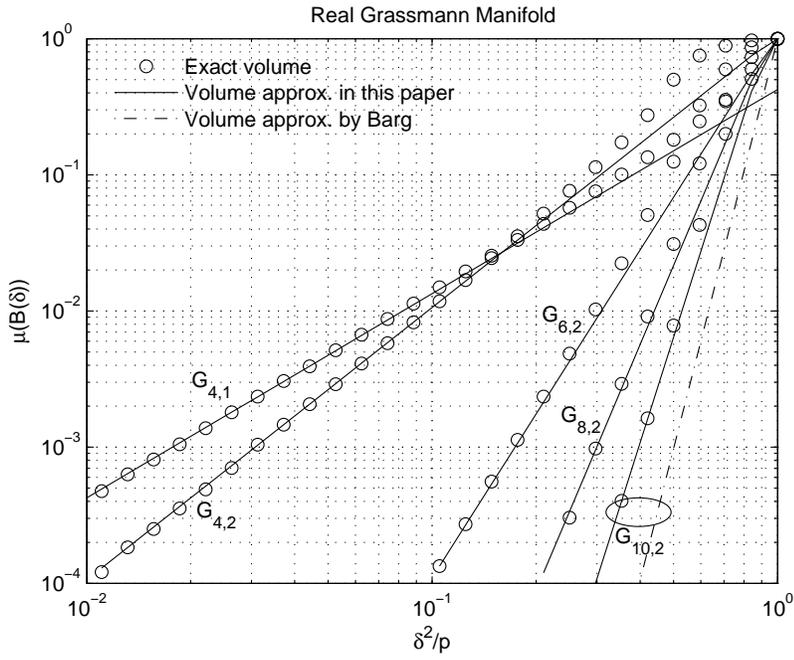}}

\subfigure[Complex Grassmann manifolds]{\includegraphics[%
  clip,
  scale=0.8]{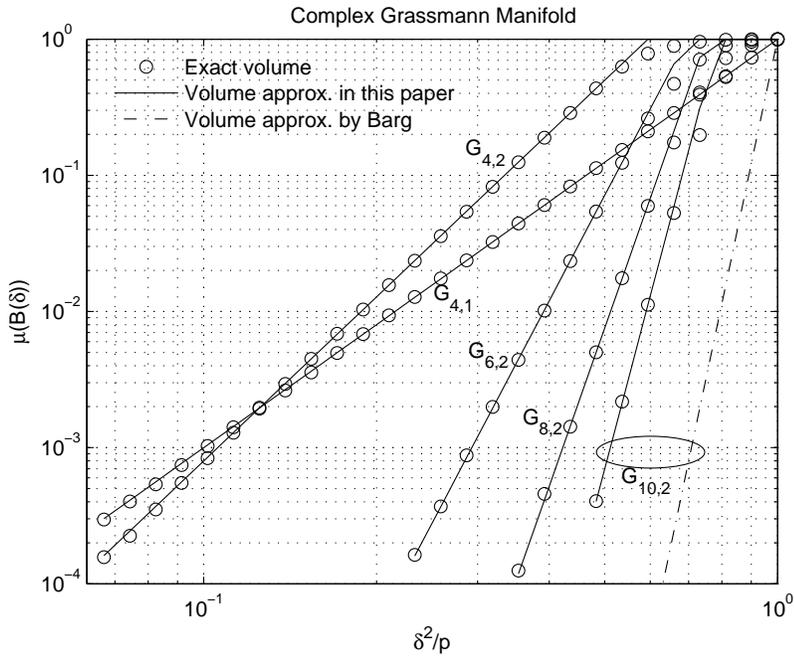}}

\caption{\label{cap:volume_in_Grassmann}The volume of a metric ball in the
Grassmann manifold}
\end{figure}

For engineering purposes, it may be satisfactory to approximate the
volume of a metric ball $B\left(\delta\right)$ by $c_{n,p,q,\beta}\delta^{\beta p\left(n-q\right)}$
when the radius $\delta$ is relatively small. Fig. \ref{cap:volume_in_Grassmann}
compares the exact volume of a metric ball (\ref{eq:actual_volume})
and the volume approximation $c_{n,p,q,\beta}\delta^{\beta p\left(n-q\right)}$.
In the simulations, we always assume $p=q$. The volume approximation
becomes exact for the complex Grassmann manifold when $\delta\leq1$.
To calculate the exact volume without appealing to Theorem \ref{thm:Volume_formula},
Monte Carlo simulation is employed to evaluate the complicated integrals
in (\ref{eq:actual_volume}). Since \[
\mu\left(B\left(\delta\right)\right)=\Pr\left\{ Q:\; d_{c}\left(P,Q\right)\leq\delta\right\} \]
 where $P\in\mathcal{G}_{n,p}\left(\mathbb{L}\right)$ is chosen arbitrarily
and $Q$ is uniformly distributed in the $\mathcal{G}_{n,p}\left(\mathbb{L}\right)$,
simulating the event $\left\{ Q:\; d_{c}\left(P,Q\right)\leq\delta\right\} $
gives $\mu\left(B\left(\delta\right)\right)$. The simulation results
for the real and complex Grassmann manifolds are presented in Fig.
\ref{cap:volume_in_Grassmann}(a) and \ref{cap:volume_in_Grassmann}(b)
respectively. Simulations show that the volume approximation (solid
lines) is close to the exact volume (circles) when the radius of the
metric ball is not large. We also compare our approximation with Barg's
approximation $\left(\delta/\sqrt{p}\right)^{\beta np}$ (from (\ref{eq:Barg-formula}))
for the case $n=10$ and $p=2$. Simulations show that the exact volume
and Barg's approximation (dash-dot lines) may not be of the same order
while the approximation in this paper is much more accurate.

\section{Quantization Bounds\label{sec:Quantization-Bounds}}

This section derives the sphere packing bounds and quantifies the
rate distortion tradeoff for both quantization problem and plane matching
problem. The results developed hold for Grassmann manifolds with arbitrary
$n$, $p$ and $q$.

\subsection{\label{sub:Sphere-Packing-Bounds}Sphere Packing Bounds}

The Gilbert-Varshamov and Hamming bounds for $\mathcal{G}_{n,p}\left(\mathbb{L}\right)$
are given in the following corollary.

\begin{cor}
\label{cor:packing_bounds}When $\delta$ is sufficiently small ($\delta\leq1$
necessarily), there exists a code $\mathcal{C}$ in the $\mathcal{G}_{n,p}\left(\mathbb{L}\right)$
with size $K$ and the minimum distance $\delta$ such that \[
c_{n,p,p,\beta}^{-1}\delta^{-\beta p\left(n-p\right)}\left(1+O\left(\delta^{2}\right)\right)\leq K.\]
For any code with the minimum distance $\delta$, \[
K\leq c_{n,p,p,\beta}^{-1}\left(\frac{\delta}{2}\right)^{-\beta p\left(n-p\right)}\left(1+O\left(\delta^{2}\right)\right).\]

\end{cor}
\begin{proof}
The corollary is proved by substituting the volume formula (\ref{eq:simplified_volume_formula})
into (\ref{eq:GV_bd}) and (\ref{eq:Hamming_bd}).
\end{proof}
\begin{remrk}
Applying Corollary \ref{cor:volume-bounds} would provide sharper
information on the higher order term. But we omit this.
\end{remrk}
%Corollary

\subsection{\label{sub:Quantization-tradeoff}Quantization: the Rate Distortion
Tradeoff}

The rate distortion tradeoff for quantization is characterized in
this subsection%
\footnote{For compositional clarity, the results for plane matching is summarized
in a separate subsection \ref{sub:Plane-Matching-Tradeoff}.%
}. Here, we assume that the quantization is on $\mathcal{G}_{n,p}\left(\mathbb{L}\right)$,
the source is uniformly distributed in $\mathcal{G}_{n,p}\left(\mathbb{L}\right)$
and the distortion metric is defined as the square of the chordal
distance. The derivation is based on the volume formula in (\ref{eq:simplified_volume_formula}).
A lower bound and an upper bound on the distortion rate function are
established. Denote the size of the code by $K$. Then the lower and
upper bounds are asymptotically identical when $p$ is fixed, $n$
and the code rate $\log_{2}K$ approach to infinity with a fixed ratio.
Therefore, these bounds precisely quantify the asymptotic rate distortion
tradeoff. Note that the upper bound is the average distortion of random
codes. Random codes are asymptotically optimal.

The following theorem gives a lower bound and an upper bound on the
distortion rate function.

\begin{thm}
\label{thm:DRF_bounds}When $K$ is sufficiently large ($\left(c_{n,p,p,\beta}K\right)^{-\frac{2}{\beta p\left(n-p\right)}}\leq1$
necessarily), the distortion rate function is bounded as in \begin{align}
 & \frac{\beta p\left(n-p\right)}{\beta p\left(n-p\right)+2}\left(c_{n,p,p,\beta}K\right)^{-\frac{2}{\beta p\left(n-p\right)}}\left(1+o\left(1\right)\right)\leq D^{*}\left(K\right)\nonumber \\
 & \quad\quad\quad\leq\frac{2\Gamma\left(\frac{2}{\beta p\left(n-p\right)}\right)}{\beta p\left(n-p\right)}\left(c_{n,p,p,\beta}K\right)^{-\frac{2}{\beta p\left(n-p\right)}}\left(1+o\left(1\right)\right)\label{eq:DRF_bounds}\end{align}

\end{thm}
\begin{remrk}
For engineering purposes, the main order terms in (\ref{eq:DRF_bounds})
are usually accurate enough to characterize the distortion rate function.
The details of the $\left(1+o\left(1\right)\right)$ correction are
spelled out in Theorem \ref{thm:DRF_bounds_details} below, where
a quantization is viewed as a specific case of plane matching.
\end{remrk}

The lower bound and the upper bound are proved in Appendix \ref{sub:Proof-of-DCF-lb}
and \ref{sub:Proof-of-DCF-ub} respectively. We sketch the proof as
follows.

The lower bound is proved by a sphere packing argument. The key is
to construct an ideal quantizer, which may not exist, to minimize
the distortion. Suppose that there exists $K$ metric balls of the
same radius $\delta_{0}$ packing and covering the whole $\mathcal{G}_{n,p}\left(\mathbb{L}\right)$
at the same time. Then the quantizer which maps each of those balls
into its center gives the minimum distortion among all quantizers.
Of course such an ideal packing  may not exist. It provides a lower
bound on the distortion rate function.

The basic idea behind the upper bound is that the distortion of any
particular code is an upper bound of the distortion rate function
and so is the average distortion of an ensemble of codes. Toward the
proof, the ensemble of random codes $\mathcal{C}_{\mathrm{rand}}=\left\{ P_{1},\cdots,P_{K}\right\} $
are employed, where the codewords $P_{i}$'s are independently drawn
from the uniform distribution on $\mathcal{G}_{n,p}\left(\mathbb{L}\right)$.
For any given code $\mathcal{C}_{\mathrm{rand}}$, the corresponding
distortion is given by \[
D\left(\mathcal{C}_{\mathrm{rand}}\right)=\mathrm{E}_{Q}\left[\underset{P_{i}\in\mathcal{C}_{\mathrm{rand}}}{\min}\; d_{c}^{2}\left(P_{i},Q\right)\right],\]
where $Q\in\mathcal{G}_{n,p}\left(\mathbb{L}\right)$ is a uniformly
distributed plane. It is clear that $D^{*}\left(K\right)\leq\mathrm{E}_{\mathcal{C}_{\mathrm{rand}}}\left[D\left(\mathcal{C}_{\mathrm{rand}}\right)\right]$.
We want to calculate $\mathrm{E}_{\mathcal{C}_{\mathrm{rand}}}\left[D\left(\mathcal{C}_{\mathrm{rand}}\right)\right]$.
Note that \begin{eqnarray*}
\mathrm{E}_{\mathcal{C}_{\mathrm{rand}}}\left[D\left(\mathcal{C}_{\mathrm{rand}}\right)\right] & = & \mathrm{E}_{\mathcal{C}_{\mathrm{rand}}}\left[\mathrm{E}_{Q}\left[\underset{P_{i}\in\mathcal{C}_{\mathrm{rand}}}{\min}\; d_{c}^{2}\left(P_{i},Q\right)\right]\right]\\
 & = & \mathrm{E}_{Q}\left[\mathrm{E}_{\mathcal{C}_{\mathrm{rand}}}\left[\underset{P_{i}\in\mathcal{C}_{\mathrm{rand}}}{\min}\; d_{c}^{2}\left(P_{i},Q\right)\right]\right].\end{eqnarray*}
Since $\mathcal{C}_{\mathrm{rand}}$ is randomly generated from the
uniform distribution, $\mathrm{E}_{\mathcal{C}_{\mathrm{rand}}}\left[\min\; d_{c}^{2}\left(P_{i},Q\right)\right]$
should be independent of the choice of $Q$. Therefore, \[
\mathrm{E}_{\mathcal{C}_{\mathrm{rand}}}\left[D\left(\mathcal{C}_{\mathrm{rand}}\right)\right]=\mathrm{E}_{\mathcal{C}_{\mathrm{rand}}}\left[\min\; d_{c}^{2}\left(P_{i},Q\right)\right]\]
for any fixed $Q$. By the volume formula and the extreme order statistics,
we are able to calculate the distribution of $d_{c}^{2}\left(P_{i},Q\right)$
(for $\forall i$) and $\min\; d_{c}^{2}\left(P_{i},Q\right)$. In
appendix \ref{sub:Proof-of-DCF-ub}, we prove that for any given $Q\in\mathcal{G}_{n,p}\left(\mathbb{L}\right)$,
$K^{\frac{2}{t}}\cdot\mathrm{E}_{\mathcal{C}_{\mathrm{rand}}}\left[D\left(\mathcal{C}_{\mathrm{rand}}\right)\right]$
converges to a constant as $K$ approaches infinity. Therefore, an
upper bound of the distortion rate function is obtained for asymptotically
large $K$. 

The rate distortion function is directly related to the distortion
rate function. The following corollary quantifies the rate distortion
function.

\begin{cor}
\label{cor:RDF_bounds}When the required distortion $D$ is sufficiently
small ($D\leq1$ necessarily), the rate distortion function satisfies
the following bounds,\begin{align}
 & \frac{1}{c_{n,p,p,\beta}}\left(\frac{\beta p\left(n-p\right)}{2\Gamma\left(\frac{2}{\beta p\left(n-p\right)}\right)}D\right)^{-\frac{\beta p\left(n-p\right)}{2}}\left(1+o\left(1\right)\right)\leq K^{*}\left(D\right)\nonumber \\
 & \quad\quad\quad\leq\frac{1}{c_{n,p,p,\beta}}\left(\frac{\beta p\left(n-p\right)+2}{\beta p\left(n-p\right)}D\right)^{-\frac{\beta p\left(n-p\right)}{2}}\left(1+o\left(1\right)\right).\label{eq:RDF_bounds}\end{align}

\end{cor}

To investigate the difference between the lower and upper bounds in
(\ref{eq:DRF_bounds}), proceed as follows. Since the exponential
terms are the same in both bounds, focus on the coefficients. The
difference between the two bounds depends on the number of real dimensions
$\beta p\left(n-p\right)$ of the underlying Grassmann manifold. There
are three cases to consider. 

\begin{case}
\label{cas:t=3D0}$\beta p\left(n-p\right)=0$. This only occurs $n=p$.
Then the whole $\mathcal{G}_{n,n}\left(\mathbb{L}\right)$ contains
only one element and no quantization is needed essentially.
\end{case}

\begin{case}
\label{cas:t=3D1}$\beta p\left(n-p\right)=1$. This happens if and
only if $\mathbb{L}=\mathbb{R}$, $n=2$ and $p=1$. In this case,
it can be verified that the principle angle $\theta$ between a uniformly
distributed $Q\in\mathcal{G}_{2,1}\left(\mathbb{R}\right)$ and any
fixed $P\in\mathcal{G}_{2,1}\left(\mathbb{R}\right)$ is uniformly
distributed in $\left[0,\frac{\pi}{2}\right]$. From here, the optimal
quantization can be explicitly constructed. Since there exists $K$
metric balls with radius $\sin\frac{\pi}{4K}$ such that those balls
not only pack but also cover the whole $\mathcal{G}_{2,1}\left(\mathbb{R}\right)$,
the quantizer mapping those balls into its center is optimal. The
distortion rate function can be explicitly calculated as\[
D^{*}\left(K\right)=\frac{1}{2K}-\frac{1}{\pi}\sin\frac{\pi}{2K}.\]

\end{case}

\begin{case}
\label{cas:t>1}$\beta p\left(n-p\right)\geq2$. For this general
case, an elementary calculation shows that \[
\frac{1}{2}\leq\frac{\beta p\left(n-p\right)}{\beta p\left(n-p\right)+2}\leq\frac{2}{\beta p\left(n-p\right)}\Gamma\left(\frac{2}{\beta p\left(n-p\right)}\right)\leq1,\]
and we expect the difference between the two bounds to decrease as
$n$ approaches infinity. Indeed, the following corollary shows that
the lower and upper bounds are asymptotically the same.
\end{case}
\begin{cor}
\label{cor:Asymptotic-Meet}Suppose that $p$ is fixed, $n$ and the
code rate $\log_{2}K$ approach to infinity simultaneously with $\bar{r}\triangleq\underset{\left(n,K\right)\rightarrow+\infty}{\lim}\frac{\log_{2}K}{n}$.
If the normalized code rate $\bar{r}$ is sufficiently large ($p2^{-\frac{2}{\beta p}\bar{r}}\leq1$
necessarily), then \[
\underset{\left(n,K\right)\rightarrow+\infty}{\lim}D^{*}\left(K\right)=p2^{-\frac{2}{\beta p}\bar{r}}.\]
On the other hand, if the required distortion $D$ is sufficiently
small ($D\leq1$ necessarily), then the minimum code size required
to achieve that distortion satisfies \[
\underset{\left(n,K\right)\rightarrow+\infty}{\lim}\frac{\log_{2}K^{*}\left(D\right)}{n}=\frac{\beta p}{2}\log_{2}\left(\frac{p}{D}\right).\]

\end{cor}
\begin{proof}
The leading order is read off from \[
\underset{n\rightarrow+\infty}{\lim}\frac{\beta p\left(n-p\right)}{\beta p\left(n-p\right)+2}=1=\underset{n\rightarrow+\infty}{\lim}\frac{2}{\beta p\left(n-p\right)}\Gamma\left(\frac{2}{\beta p\left(n-p\right)}\right),\]
\[
\underset{n\rightarrow+\infty}{\lim}\left(c_{n,p,p,\beta}\right)^{-\frac{2}{\beta p\left(n-p\right)}}=p,\]
\[
\underset{\left(n,K\right)\rightarrow+\infty}{\lim}K^{-\frac{2}{p\left(n-p\right)}}=2^{-\frac{2}{\beta p}\bar{r}},\]
and \[
Kc_{n,p,p,\beta}\overset{\left(n,K\right)\rightarrow\infty}{\longrightarrow}\infty.\]
That the $\left(1+o\left(1\right)\right)$ multiplicative errors fall
into place is the content of Theorem \ref{thm:DRF_bounds_details}.
\end{proof}

The lower and upper bounds asymptotically agreeing accurately quantifies
the distortion rate function. Since the upper bound is actually derived
from the average distortion of random codes, this implies that \emph{random
codes are asymptotically optimal}.

\begin{figure}
\includegraphics[%
  clip,
  scale=0.8]{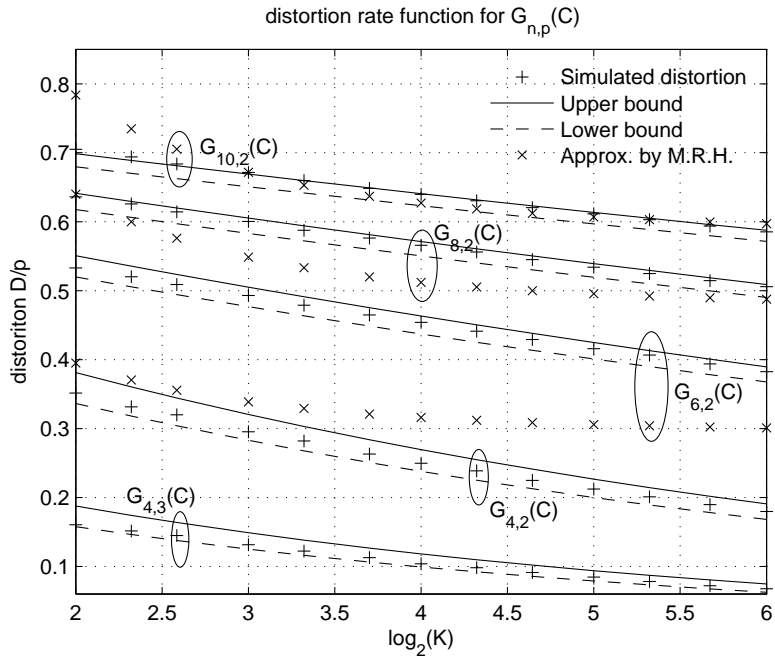}

\caption{\label{cap:DRF_bounds}Bounds on the distortion rate function}
\end{figure}

As a comparison, we cite the distortion rate function approximation
derived in \cite{Heath_ICASSP05_Quantization_Grassmann_Manifold}.
For $\mathcal{G}_{n,p}\left(\mathbb{C}\right)$ with $p=1$, that
paper offers the approximation  \begin{equation}
D^{*}\left(K\right)\approx\left(\frac{n-1}{n}\right)K^{-\frac{1}{n-1}}\label{eq:Heath_approx_p1}\end{equation}
 by asymptotic arguments. According to our results in Theorem \ref{thm:DRF_bounds},
the approximation (\ref{eq:Heath_approx_p1}) is indeed a lower bound
for the distortion rate function and valid for all possible $n$'s.
For $\mathcal{G}_{n,p}\left(\mathbb{C}\right)$ with $p$ fixed and
$n\gg p$, a lower bound of an upper bound on the distortion rate
function is given in \cite{Heath_ICASSP05_Quantization_Grassmann_Manifold}
based on an estimation of the minimum distance of a code. It is less
robust than the result in Theorem \ref{thm:DRF_bounds} in that it
is neither a lower bound nor an upper bound, and it only holds for
$n\gg p$ (see Fig. \ref{cap:DRF_bounds} for an empirical comparison). 

Besides characterizing the rate distortion tradeoff, we are also interested
in designing a code to minimize distortion for a given code size $K$.
Generally speaking, it is computational complicated to design a code
to minimize distortion directly. In \cite{Love_IT03_Grassman_Beamforming_MIMO}
and \cite{Love_SP05_Limited_feedback_unitary_precoding}, a suboptimal
design criterion, i.e., maximization the minimum distance between
codeword pairs, is proposed to reduce computational complexity. Refer
this suboptimal criterion as \emph{max-min} criterion. According to
our volume formula (\ref{eq:simplified_volume_formula}), the same
criterion can be verified. Let the minimum distance of a code $\mathcal{C}$
be $\delta$. Note that the metric balls of radius $\frac{\delta}{2}$
and centered at $P_{i}\in\mathcal{C}$ are disjoint. Then the corresponding
distortion is upper bounded by \begin{equation}
D\left(\mathcal{C}\right)\leq\frac{\delta^{2}}{4}K\mu\left(B\left(\delta/2\right)\right)+p\left(1-K\mu\left(B\left(\delta/2\right)\right)\right).\label{eq:Distortion_minimum_distance}\end{equation}
Apply the volume formula (\ref{eq:simplified_volume_formula}). An
elementary calculation shows that the first derivative of the upper
bound is negative when \[
\delta<\sqrt{\frac{4\beta p^{2}\left(n-p\right)}{2+\beta p\left(n-p\right)}}.\]
This property implies the upper bound (\ref{eq:Distortion_minimum_distance})
is a decreasing function of $\delta$ when $\delta$ is small enough.
Thus, max-min criterion is an appropriate design criterion to obtain
codes with small distortion. Since this criterion only requires to
calculate the distance between codeword pairs, the computational complexity
is less than that of designing a code to minimize the distortion directly.

Fig. \ref{cap:DRF_bounds} compares the simulated distortion rate
function (the plus markers) with its lower bound (the dashed lines)
and upper bound (the solid lines) in (\ref{eq:DRF_bounds}). To simulate
the distortion rate function, we use the max-min criterion to design
codes and use the minimum distortion of the designed codes as the
distortion rate function. Simulation results show that the bounds
in (\ref{eq:DRF_bounds}) hold for large $K$. When $K$ is relatively
small, the formula (\ref{eq:DRF_bounds}) can serve as good approximations
to the distortion rate function as well. Simulations also verify the
previous discussion on the difference between the two bounds. The
difference between the bounds is small and it becomes smaller as $n$
increases. In addition, we compare our bounds with the approximation
(the {}``x'' markers) derived in \cite{Heath_ICASSP05_Quantization_Grassmann_Manifold}.
Simulations show that the approximation in \cite{Heath_ICASSP05_Quantization_Grassmann_Manifold}
is neither an upper bound nor a lower bound. It works for the case
that $n=10$ and $p=2$ but doesn't work when $n\leq8$ and $p=2$.
As a comparison, the bounds (\ref{eq:DRF_bounds}) derived in this
paper hold for arbitrary $n$ and $p$.

\subsection{\label{sub:Plane-Matching-Tradeoff}Plane Matching: the Rate Distortion
Tradeoff}

For completeness, this subsection summarizes the results about the
rate distortion tradeoff for plane matching. The corresponding proofs
follow those for the quantization problem.

Let a code $\mathcal{C}\in\mathcal{G}_{n,p}\left(\mathbb{L}\right)$.
The plane matching problem is to choose a plane in $\mathcal{C}$
to match a random plane $Q\in\mathcal{G}_{n,q}\left(\mathbb{L}\right)$.
Without loss of generality, we assume $1\leq p\leq q\leq n$. Denote
the size of the code $\mathcal{C}$ by $K$. When $K$ is sufficient
large, the distortion rate function is bounded by\begin{align*}
 & \frac{\beta p\left(n-q\right)}{\beta p\left(n-q\right)+2}\left(c_{n,p,q,\beta}K\right)^{-\frac{2}{\beta p\left(n-q\right)}}\left(1+o\left(1\right)\right)\leq D^{*}\left(K\right)\\
 & \quad\quad\quad\leq\frac{2\Gamma\left(\frac{2}{\beta p\left(n-q\right)}\right)}{\beta p\left(n-q\right)}\left(c_{n,p,q,\beta}K\right)^{-\frac{2}{\beta p\left(n-q\right)}}\left(1+o\left(1\right)\right).\end{align*}
When the required distortion $D$ is sufficiently small, the rate
distortion function is bounded by\begin{align*}
 & \frac{1}{c_{n,p,q,\beta}}\left(\frac{\beta p\left(n-q\right)}{2\Gamma\left(\frac{2}{\beta p\left(n-q\right)}\right)}D\right)^{-\frac{\beta p\left(n-q\right)}{2}}\left(1+o\left(1\right)\right)\leq K^{*}\left(D\right)\\
 & \quad\quad\quad\leq\frac{1}{c_{n,p,q,\beta}}\left(\frac{\beta p\left(n-q\right)+2}{\beta p\left(n-q\right)}D\right)^{-\frac{\beta p\left(n-q\right)}{2}}\left(1+o\left(1\right)\right).\end{align*}
We finally detail the above $\left(1+o\left(1\right)\right)$ errors,
and so those in (\ref{eq:DRF_bounds}) and (\ref{eq:RDF_bounds})
as well with the following theorem.

\begin{thm}
\label{thm:DRF_bounds_details}Let $a$ be an arbitrary real number
such that $0<a<1$ and $K$ be sufficiently large ($\left(c_{n,p,q,\beta}K\right)^{-\frac{2}{\beta p\left(n-q\right)}}\leq1$
necessarily). If $\mathbb{L}=\mathbb{R}$ and $q=p$, then \begin{align*}
 & \frac{p\left(n-p\right)}{p\left(n-p\right)+2}\left(c_{n,p,p,1}K\right)^{-\frac{2}{p\left(n-p\right)}}\left(1-\left(c_{n,p,p,1}K\right)^{-\frac{2}{p\left(n-p\right)}}\right)^{\frac{1}{n-p}}\leq D^{*}\left(K\right)\\
 & \quad\quad\leq\frac{2\Gamma\left(\frac{2}{p\left(n-p\right)}\right)}{p\left(n-p\right)}\left(Kc_{n,p,p,1}\right)^{-\frac{2}{p\left(n-p\right)}}+p\exp\left(-\left(Kc_{n,p,p,1}\right)^{1-a}\right).\end{align*}
If $\mathbb{L}=\mathbb{R}$ and $q=p+1$, or $\mathbb{L}=\mathbb{C}$
and $q=p$, then \begin{align*}
 & \frac{\beta p\left(n-q\right)}{\beta p\left(n-q\right)+2}\left(c_{n,p,q,\beta}K\right)^{-\frac{2}{\beta p\left(n-q\right)}}\leq D^{*}\left(K\right)\\
 & \quad\quad\leq\frac{2\Gamma\left(\frac{2}{\beta p\left(n-q\right)}\right)}{\beta p\left(n-q\right)}\left(Kc_{n,p,q,\beta}\right)^{-\frac{2}{\beta p\left(n-q\right)}}+p\exp\left(-\left(Kc_{n,p,q,\beta}\right)^{1-a}\right).\end{align*}
If $\mathbb{L}=\mathbb{R}$ and $q>p+1$, or $\mathbb{L}=\mathbb{C}$
and $q>p$, then \begin{align*}
 & \frac{\beta p\left(n-q\right)}{\beta p\left(n-q\right)+2}\left(c_{n,p,q,\beta}K\right)^{-\frac{2}{\beta p\left(n-q\right)}}\leq D^{*}\left(K\right)\\
 & \quad\quad\leq\frac{2\Gamma\left(\frac{2}{\beta p\left(n-q\right)}\right)}{\beta p\left(n-q\right)}\left(Kc_{n,p,q,\beta}\right)^{-\frac{2}{\beta p\left(n-q\right)}}\left(1-\left(Kc_{n,p,q,\beta}\right)^{-\frac{2a}{\beta p\left(n-q\right)}}\right)^{-\frac{\beta p\left(q-p+1\right)-2p}{\beta p\left(n-q\right)}}\\
 & \quad\quad\quad+p\exp\left(-\left(Kc_{n,p,q,\beta}\right)^{1-a}\left(1-\left(Kc_{n,p,q,\beta}\right)^{-\frac{2a}{\beta p\left(n-q\right)}}\right)^{\frac{\beta p\left(q-p+1\right)}{2}-p}\right).\end{align*}

\end{thm}
\begin{proof}
The proof is given in Appendix \ref{sub:Proof-of-DCF-lb} and \ref{sub:Proof-of-DCF-ub}.
\end{proof}

The lower and upper bounds are asymptotically identical. Let $p$
and $q$ be fixed. Let $n$ and the code rate $\log_{2}K$ approach
to infinity simultaneously with $\bar{r}=\underset{\left(n,K\right)\rightarrow\infty}{\lim}\frac{\log_{2}K}{n}$.
If the normalized code rate $\bar{r}$ is sufficiently large, then
\[
\underset{\left(n,K\right)\rightarrow+\infty}{\lim}D^{*}\left(K\right)=p2^{-\frac{2}{\beta p}\bar{r}}.\]
On the other hand, if the required distortion $D$ is sufficiently
small, then the minimum code size required to achieve that distortion
satisfies \[
\underset{\left(n,K\right)\rightarrow+\infty}{\lim}\frac{\log_{2}K^{*}\left(D\right)}{n}=\frac{\beta p}{2}\log_{2}\left(\frac{p}{D}\right).\]

\section{\label{sec:Application}An Application to MIMO Systems with Finite
Rate Channel State Feedback}

As an application of the derived quantization bounds on the Grassmann
manifold, this section discusses the information theoretical benefit
of finite-rate channel-state feedback for MIMO systems using power
on/off strategy. In particular, we show that the benefit of the channel
state feedback can be accurately characterized by the distortion of
a quantization on the Grassmann manifold. 

The effect of finite-rate feedback on MIMO systems using power on/off
strategy has been widely studied. MIMO systems with only one on-beam
are discussed in \cite{Sabharwal_IT03_Beamforming_MIMO} and \cite{Love_IT03_Grassman_Beamforming_MIMO},
where the beamforming codebook design criterion and performance analysis
are derived by geometric arguments in the Grassmann manifold $\mathcal{G}_{n,1}\left(\mathbb{C}\right)$.
MIMO systems with multiple on-beams are considered in \cite{Heath_TSigProcess_Multi_mode_Antenna_Selection,Love_IT05sub_limited_feedback_unitary_precoding,Heath_ICASSP05_Quantization_Grassmann_Manifold,Rao_icc05_MIMO_spatial_multiplexing_limit_feedback,Honig_MiliComm03_Asymptotic_MIMO_Limited_Feedback}.
Criteria to select the beamforming matrix are developed in \cite{Heath_TSigProcess_Multi_mode_Antenna_Selection}
and \cite{Love_IT05sub_limited_feedback_unitary_precoding}. The signal-to-noise
ratio (SNR) loss due to quantized beamforming is discussed in \cite{Heath_ICASSP05_Quantization_Grassmann_Manifold}.
The corresponding analysis is based on Barg's formula (\ref{eq:Barg-formula})
and only valid for MIMO systems with asymptotically large number of
transmit antennas. The effect of beamforming quantization on information
rate is investigated in \cite{Rao_icc05_MIMO_spatial_multiplexing_limit_feedback}
and \cite{Honig_MiliComm03_Asymptotic_MIMO_Limited_Feedback}. The
loss in information rate is quantified for high SNR region in \cite{Rao_icc05_MIMO_spatial_multiplexing_limit_feedback}.
That analysis is based on an approximation of the logdet function
in the high SNR region and a metric on the Grassmann manifold other
than the chordal distance. In \cite{Honig_MiliComm03_Asymptotic_MIMO_Limited_Feedback},
a formula to calculate the information rate for all SNR regimes is
proposed by letting the numbers of transmit antennas, receive antennas
and feedback rate approach infinity simultaneously. But this formula
overestimates the performance in general. 

\begin{figure}
\includegraphics[%
  clip,
  scale=0.8]{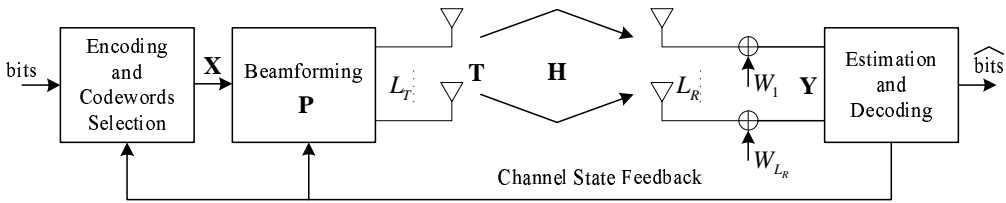}

\caption{\label{cap:System-model}System model for a MIMO system}
\end{figure}

The basic model of a wireless communication system with $L_{T}$ transmit
antennas, $L_{R}$ receive antennas and finite-rate channel state
feedback is given in Fig. \ref{cap:System-model}. The information
bit stream is encoded into the Gaussian signal vector $\mathbf{X}\in\mathbb{C}^{s\times1}$
and then multiplied by the beamforming matrix $\mathbf{P}\in\mathbb{C}^{L_{T}\times s}$
to generate the transmitted signal $\mathbf{T}=\mathbf{PX}$, where
$s$ is the dimension of the signal $\mathbf{X}$ satisfying $1\leq s\leq L_{T}$
and the beamforming matrix $\mathbf{P}$ satisfies $\mathbf{P}^{\dagger}\mathbf{P}=\mathbf{I}_{s}$.
In power on/off strategy, $\mathrm{E}\left[\mathbf{X}\mathbf{X}^{\dagger}\right]=P_{\mathrm{on}}\mathbf{I}_{s}$
where the $P_{\mathrm{on}}$ constant denotes the on-power. Assume
that the channel $\mathbf{H}$ is Rayleigh flat fading, i.e., the
entries of $\mathbf{H}$ are independent and identically distributed
(i.i.d.) circularly symmetric complex Gaussian variables with zero
mean and unit variance ($\mathcal{CN}\left(0,1\right)$) and $\mathbf{H}$
is i.i.d. for each channel use. Let $\mathbf{Y}\in\mathbb{C}^{L_{R}\times1}$
be the received signal and $\mathbf{W}\in\mathbb{C}^{L_{R}\times1}$
be the Gaussian noise, then\[
\mathbf{Y}=\mathbf{HPX}+\mathbf{W},\]
where $E\left[\mathbf{W}\mathbf{W}^{\dagger}\right]=\mathbf{I}_{L_{R}}$.
We also assume that there is a beamforming codebook $\mathcal{B}=\left\{ \mathbf{P}_{i}\in\mathbb{C}^{L_{T}\times s}:\right.$
$\left.\mathbf{P}_{i}^{\dagger}\mathbf{P}_{i}=\mathbf{I}_{s}\right\} $
declared to both the transmitter and the receiver before the transmission.
At the beginning of each channel use, the channel state $\mathbf{H}$
is perfectly estimated at the receiver. A message, which is a function
of the channel state, is sent back to the transmitter through a feedback
channel. The feedback is error-free and rate limited. According to
the channel state feedback, the transmitter chooses an appropriate
beamforming matrix $\mathbf{P}_{i}\in\mathcal{B}$. Let the feedback
rate be $R_{\mathrm{fb}}$bits/channel use. Then the size of the beamforming
codebook $\left|\mathcal{B}\right|\leq2^{R_{\mathrm{fb}}}$. The feedback
function is a mapping from the set of channel state into the beamforming
matrix index set, $\varphi:\;\left\{ \mathbf{H}\right\} \rightarrow\left\{ i:\;1\leq i\leq\left|\mathcal{B}\right|\right\} $.
This section will quantify the corresponding information rate \[
\mathcal{I}=\underset{\mathcal{B}:\left|\mathcal{B}\right|\leq2^{R_{\mathrm{fb}}}}{\max}\underset{\varphi}{\max}\;\mathrm{E}\left[\log\left|\mathbf{I}_{L_{R}}+P_{\mathrm{on}}\mathbf{H}\mathbf{P}_{\varphi\left(\mathbf{H}\right)}\mathbf{P}_{\varphi\left(\mathbf{H}\right)}^{\dagger}\mathbf{H}\right|\right],\]
where $P_{\mathrm{on}}=\rho/s$ and $\rho$ is the average received
SNR. 

Before discussing the finite-rate feedback case, we consider the case
that the transmitter has full knowledge of the channel state $\mathbf{H}$.
In this setting, the optimal beamforming matrix is given by $\mathbf{P}_{\mathrm{opt}}=\mathbf{V}_{s}$
where $\mathbf{V}_{s}\in\mathbb{C}^{L_{T}\times s}$ is the matrix
composed by the right singular vectors of $\mathbf{H}$ corresponding
to the largest $s$ singular values \cite{Dai_05_Power_onoff_strategy_design_finite_rate_feedback}.
The corresponding information rate is \begin{equation}
\mathcal{I}_{\mathrm{opt}}=\mathrm{E}_{\mathbf{H}}\left[\sum_{i=1}^{s}\mathrm{ln}\left(1+P_{\mathrm{on}}\lambda_{i}\right)\right],\label{eq:I_perfect_beamforming}\end{equation}
where $\lambda_{i}$ is the $i^{\mathrm{th}}$ largest eigenvalue
of $\mathbf{H}\mathbf{H}^{\dagger}$. In \cite[Section III]{Dai_05_Power_onoff_strategy_design_finite_rate_feedback},
we derive an asymptotic formula to approximate a quantity of the form
$\mathrm{E}_{\mathbf{H}}\left[\sum_{i=1}^{s}\ln\left(1+c\lambda_{i}\right)\right]$
where $c>0$ is a constant. Apply the asymptotic formula in \cite{Dai_05_Power_onoff_strategy_design_finite_rate_feedback}.
$\mathcal{I}_{\mathrm{opt}}$ can be well approximated.

The effect of finite-rate feedback can be characterized by the quantization
bounds in the Grassmann manifold. For finite-rate feedback, we define
a suboptimal feedback function\begin{equation}
i=\varphi\left(\mathbf{H}\right)\triangleq\underset{1\leq i\leq\left|\mathcal{B}\right|}{\arg\ \min}\; d_{c}^{2}\left(\mathcal{P}\left(\mathbf{P}_{i}\right),\mathcal{P}\left(\mathbf{V}_{s}\right)\right),\label{eq:feedback-fn-suboptimal}\end{equation}
where $\mathcal{P}\left(\mathbf{P}_{i}\right)$ and $\mathcal{P}\left(\mathbf{V}_{s}\right)$
are the planes in the $\mathcal{G}_{L_{T},s}\left(\mathbb{C}\right)$
generated by $\mathbf{P}_{i}$ and $\mathbf{V}_{s}$ respectively.
In \cite{Dai_05_Power_onoff_strategy_design_finite_rate_feedback},
we showed that this feedback function is asymptotically optimal as
$R_{\mathrm{fb}}\rightarrow+\infty$ and near optimal when $R_{\mathrm{fb}}<+\infty$.
With this feedback function and assuming that the feedback rate $R_{\mathrm{fb}}$
is large, it has also been shown in \cite{Dai_05_Power_onoff_strategy_design_finite_rate_feedback}
that\begin{eqnarray}
\mathcal{I} & \approx & \mathrm{E}_{\mathbf{H}}\left[\sum_{i=1}^{s}\ln\left(1+\eta_{\sup}P_{\mathrm{on}}\lambda_{i}\right)\right],\label{eq:I_finite_feedback}\end{eqnarray}
where \begin{eqnarray}
\eta_{\sup} & \triangleq & 1-\frac{1}{s}\underset{\mathcal{B}:\left|\mathcal{B}\right|\leq2^{R_{\mathrm{fb}}}}{\inf}\;\mathrm{E}_{\mathbf{V}_{s}}\left[\underset{1\leq i\leq\left|\mathcal{B}\right|}{\ \min}\; d_{c}^{2}\left(\mathcal{P}\left(\mathbf{P}_{i}\right),\mathcal{P}\left(\mathbf{V}_{s}\right)\right)\right]\nonumber \\
 & = & 1-\frac{1}{s}D^{*}\left(2^{R_{\mathrm{fb}}}\right).\label{eq:PEF-DRF}\end{eqnarray}
Thus, the difference between perfect beamforming case (\ref{eq:I_perfect_beamforming})
and finite-rate feedback case (\ref{eq:I_finite_feedback}) is quantified
by $\eta_{\sup}$, which depends on the distortion rate function on
the $\mathcal{G}_{L_{T},s}\left(\mathbb{C}\right)$. Substituting
quantization bounds (\ref{eq:DRF_bounds}) into (\ref{eq:PEF-DRF})
and applying the asymptotic formula in \cite{Dai_05_Power_onoff_strategy_design_finite_rate_feedback}
for $\mathrm{E}_{\mathbf{H}}\left[\sum_{i=1}^{s}\ln\left(1+c\lambda_{i}\right)\right]$
produce approximations of the information rate $\mathcal{I}$ as a
function of the feedback rate $R_{\mathrm{fb}}$. 

\begin{figure}
\includegraphics[%
  clip,
  scale=0.8]{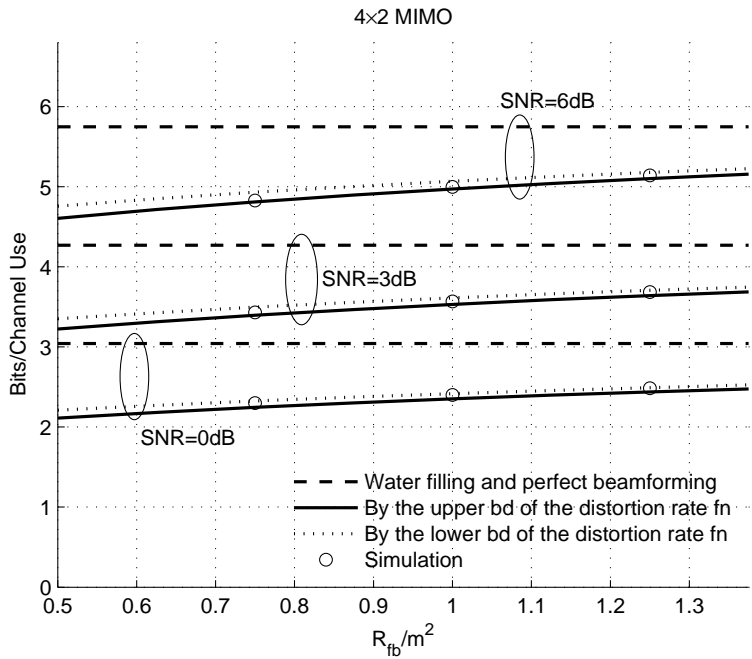}

\caption{\label{cap:Fig-Honig}Performance of a constant number of on-beams
v.s. feedback rate}
\end{figure}

\begin{figure}
\subfigure[Average mutual information]{\includegraphics[%
  clip,
  scale=0.8]{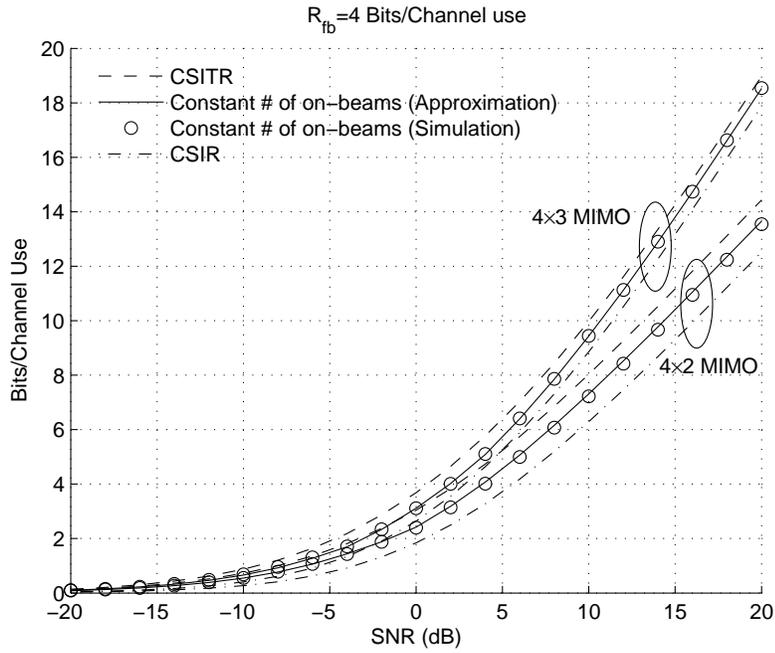}}

\subfigure[Relative performance]{\includegraphics[%
  clip,
  scale=0.8]{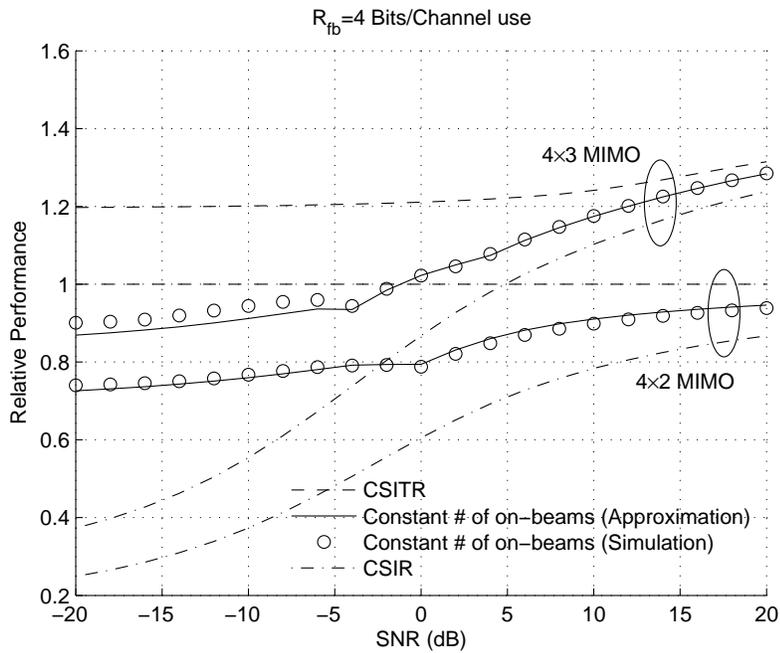}}

\caption{\label{cap:Fig_performance_comparison}Performance of a constant
number of on-beams v.s. SNR}
\end{figure}

Simulations verify the above approximations. Let $m=\min\left(L_{T},L_{R}\right)$.
Fig. \ref{cap:Fig-Honig} compares the simulated information rate
(circles) and approximations as functions of $R_{\mathrm{fb}}/m^{2}$.
The information rate approximated by the lower bound (solid lines)
and the upper bound (dotted lines) in (\ref{eq:DRF_bounds}) are presented.
The simulation results show that the performances approximated by
the bounds (\ref{eq:DRF_bounds}) match the actual performance almost
perfectly. As a comparison, the approximation proposed in \cite{Honig_MiliComm03_Asymptotic_MIMO_Limited_Feedback,Honig_Allerton03_Benefits_Limited_Feedback_Wireless_Channels},
which is based on asymptotic analysis and Gaussian approximation,
overestimates the information rate. Furthermore, we compare the simulated
information rate and the approximations for a large range of SNRs
in Fig. \ref{cap:Fig_performance_comparison}. Without loss of generality,
we only present the lower bound in (\ref{eq:DRF_bounds}) because
it corresponds to the random codes and can be achieved by appropriate
code design. Fig. \ref{cap:Fig_performance_comparison}(a) shows that
the difference between the simulated and approximated information
rate is almost unnoticeable. To make the performance difference clearer,
Fig. \ref{cap:Fig_performance_comparison}(b) gives the relative performance
as the ratio of the considered performance and the capacity of a $4\times2$
MIMO achieved by water filling power control. The difference in relative
performance is also small for all SNR regimes.

\section{Conclusion\label{sec:Conclusion}}

This paper considers the quantization problem on the Grassmann manifold.
Based on an explicit volume formula for a metric ball in the $\mathcal{G}_{n,p}\left(\mathbb{L}\right)$,
sphere packing bounds are obtained and the rate distortion tradeoff
is accurately characterized by establishing bounds on the distortion
function. Simulations verify the developed results. As an application
of the derived quantization bounds, the information rate of a MIMO
system with finite-rate channel-state feedback and power on/off strategy
is accurately quantified for the first time.

\appendix

\subsection{\label{sub:Proof-of-Volume-Formula}Proof of Theorem \ref{thm:Volume_formula}}

The proof is divided into three parts, in which we calculate the volume
formula for the $1\leq p\leq q\leq\frac{n}{2}$, $\frac{n}{2}\leq p\leq q\leq n$
and $1\leq p\leq\frac{n}{2}\leq q\leq n$ cases respectively.

$ $

\subsubsection{$1\leq p\leq q\leq\frac{n}{2}$ case}

First we prove the basic form \[
\mu\left(B\left(\delta\right)\right)=c_{n,p,q,\beta}\delta^{\beta p\left(n-q\right)}\left(1+c_{n,p,q,\beta}^{\left(1\right)}\delta^{2}+o\left(\delta^{2}\right)\right)\]
for $\delta\leq1$. Afterward, we calculate the constants $c_{n,p,q,\beta}$
and $c_{n,p,q,\beta}^{\left(1\right)}$. 

The volume of a metric ball $\mu\left(B\left(\delta\right)\right)$
is given by \begin{equation}
\mu\left(B\left(\delta\right)\right)=\underset{\sqrt{\sum_{i=1}^{p}\sin^{2}\theta_{i}}\leq\delta}{\int\cdots\int}\; d\mu_{\bm{\theta}},\label{eq:volume-formula}\end{equation}
where the differential form $d\mu_{\bm{\theta}}$ is the joint density
of $\theta_{i}$'s. For convenience, we introduce the following notations.
Define $r_{i}\triangleq\cos\theta_{i}$ and order $r_{i}$'s such
that $r_{i}\leq r_{j}$ ($\theta_{i}\geq\theta_{j}$) if $i<j$. Define
$\mathbf{r}=\left[r_{1},\cdots,r_{p}\right]$ and also \[
\left|\Delta_{p}\left(\mathbf{r}^{2}\right)\right|=\prod_{i<j}^{p}\left|r_{j}^{2}-r_{i}^{2}\right|.\]
Recall that $\beta=1$ for $\mathbb{L}=\mathbb{R}$ and $\beta=2$
for $\mathbb{L}=\mathbb{C}$. With these notations, the invariant
measure $d\mu_{\bm{\theta}}$ can be written as follows \cite{Adler_2001_Integrals_Grassmann}.\begin{equation}
d\mu_{\bm{\theta}}=v_{n,p,q,\beta}\left|\Delta_{p}\left(\mathbf{r}^{2}\right)\right|^{\beta}\prod_{i=1}^{p}\left(\left(r_{i}^{2}\right)^{\frac{\beta}{2}\left(q-p+1\right)-1}\left(1-r_{i}^{2}\right)^{\frac{\beta}{2}\left(n-p-q+1\right)-1}dr_{i}^{2}\right),\label{eq:d_mu_0}\end{equation}
where the constant $v_{n,p,q,\beta}$ is given by \begin{equation}
v_{n,p,q,\beta}=\prod_{i=1}^{p}\frac{\Gamma\left(1+\frac{\beta}{2}\right)\Gamma\left(\frac{\beta}{2}\left(n-i+1\right)\right)}{\Gamma\left(\frac{\beta}{2}i+1\right)\Gamma\left(\frac{\beta}{2}\left(n-p-i+1\right)\right)\Gamma\left(\frac{\beta}{2}\left(q-i+1\right)\right)}.\label{eq:v_n_p_q_beta}\end{equation}

To get the form (\ref{eq:simplified_volume_formula}), we perform
the variable change $\delta^{2}x_{i}=1-r_{i}^{2}$ for $1\leq i\leq p$.
Under this transformation, the integral domain \[
D_{\delta}\triangleq\left\{ \mathbf{r}:\;\sum\left(1-r_{i}^{2}\right)\leq\delta^{2},\;0\leq r_{i}^{2}\leq1\;\mathrm{for}\; i=1,\cdots,p\right\} \]
 is changed to \begin{eqnarray*}
D_{1} & \triangleq & \left\{ \mathbf{x}:\;\sum x_{i}\leq1,\;0\leq x_{i}\leq\frac{1}{\delta^{2}}\;\mathrm{for}\; i=1,\cdots,p\right\} \\
 & = & \left\{ \mathbf{x}:\;\sum x_{i}\leq1,\;0\leq x_{i}\;\mathrm{for}\; i=1,\cdots,p\right\} ,\end{eqnarray*}
where the last equation holds since $\delta\leq1$. Thus \[
\mu\left(B\left(\delta\right)\right)=v_{n,p,q,\beta}\delta^{\beta p\left(n-q\right)}\underset{D_{1}}{\int\cdots\int}\left|\Delta\left(\mathbf{x}\right)\right|^{\beta}\prod_{i=1}^{p}\left(\left(1-\delta^{2}x_{i}\right)^{\frac{\beta}{2}\left(q-p+1\right)-1}\left(x_{i}\right)^{\frac{\beta}{2}\left(n-p-q+1\right)-1}dx_{i}\right).\]
Next note that \[
\left(1-\delta^{2}x_{i}\right)^{\frac{\beta}{2}\left(q-p+1\right)-1}=1-\left(\frac{\beta}{2}\left(q-p+1\right)-1\right)x_{i}\delta^{2}+o\left(\delta^{2}\right),\]
and so we are able to express the volume of $B\left(\delta\right)$
in the desired form with \[
c_{n,p,q,\beta}\triangleq v_{n,p,q,\beta}\underset{D_{1}}{\int\cdots\int}\left|\Delta\left(\mathbf{x}\right)\right|^{\beta}\prod_{i=1}^{p}\left(\left(x_{i}\right)^{\frac{\beta}{2}\left(n-p-q+1\right)-1}dx_{i}\right),\]
and \[
c_{n,p,q,\beta}^{\left(1\right)}\triangleq-p\cdot\left(\frac{\beta}{2}\left(q-p+1\right)-1\right)\frac{\underset{D_{1}}{\int\cdots\int}x_{1}\left|\Delta\left(\mathbf{x}\right)\right|^{\beta}\prod_{i=1}^{p}\left(\left(x_{i}\right)^{\frac{\beta}{2}\left(n-p-q+1\right)-1}dx_{i}\right)}{\underset{D_{1}}{\int\cdots\int}\left|\Delta\left(\mathbf{x}\right)\right|^{\beta}\prod_{i=1}^{p}\left(\left(x_{i}\right)^{\frac{\beta}{2}\left(n-p-q+1\right)-1}dx_{i}\right)}.\]

In order to calculate the constants $c_{n,p,q,\beta}$ and $c_{n,p,q,\beta}^{\left(1\right)}$,
we need the following lemma \cite{Mehta_book2004_random_matrices}.

\begin{lemma}
\label{lem:2nd-beta-integral}It holds that \begin{eqnarray*}
 &  & \int\cdots\int x_{1}\cdots x_{m}\left|\Delta\left(\mathbf{x}\right)\right|^{\beta}\left(1-\sum_{i=1}^{p}x_{i}\right)^{\gamma-1}\prod_{i=1}^{p}x_{i}^{\alpha-1}dx_{i}\\
 &  & =\frac{\Gamma\left(\gamma\right)}{\Gamma\left(\gamma+m+\alpha p+\frac{\beta}{2}p\left(p-1\right)\right)}\prod_{i=1}^{m}\left(\alpha+\frac{\beta}{2}\left(p-i\right)\right)\\
 &  & \quad\times\prod_{i=1}^{p}\frac{\Gamma\left(\alpha+\frac{\beta}{2}\left(p-i\right)\right)\Gamma\left(1+\frac{\beta}{2}i\right)}{\Gamma\left(1+\frac{\beta}{2}\right)}.\end{eqnarray*}
where $0\leq m\leq p$, $\Re\left(\alpha\right)>0$, $\Re\left(\gamma\right)>0$,
$\Re\left(\frac{\beta}{2}\right)>-\min\left(\frac{1}{p},\frac{\Re\left(\alpha\right)}{p-1},\frac{\Re\left(\gamma\right)}{p-1}\right)$
and the integral is taken over $0\leq x_{i},\;\sum_{i=1}^{p}x_{i}\leq1$. 
\end{lemma}
\begin{proof}
This is Selberg's second generalization of the beta integral. See
\cite[Section 17.10]{Mehta_book2004_random_matrices} for a detailed
proof.
\end{proof}

According to Lemma \ref{lem:2nd-beta-integral}, we have that \begin{eqnarray*}
 &  & \underset{D_{1}}{\int\cdots\int}\left|\Delta\left(\mathbf{x}\right)\right|^{\beta}\prod_{i=1}^{p}\left(\left(x_{i}\right)^{\frac{\beta}{2}\left(n-p-q+1\right)-1}dx_{i}\right)\\
 &  & =\frac{1}{\Gamma\left(p\frac{\beta}{2}\left(n-p-q+1\right)+\frac{\beta}{2}p\left(p-1\right)+1\right)}\\
 &  & \quad\times\prod_{i=1}^{p}\frac{\Gamma\left(1+\frac{\beta}{2}i\right)\Gamma\left(\frac{\beta}{2}\left(n-p-q+1\right)+\frac{\beta}{2}\left(p-i\right)\right)}{\Gamma\left(1+\frac{\beta}{2}\right)}\\
 &  & =\frac{1}{\Gamma\left(\frac{\beta}{2}p\left(n-q\right)+1\right)}\prod_{i=1}^{p}\frac{\Gamma\left(1+\frac{\beta}{2}i\right)\Gamma\left(\frac{\beta}{2}\left(n-q-i+1\right)\right)}{\Gamma\left(1+\frac{\beta}{2}\right)}.\end{eqnarray*}
Substituting the formula (\ref{eq:v_n_p_q_beta}) for $v_{n,p,q,\beta}$
into the integral expansion of $c_{n,p,q,\beta}$ yields\[
c_{n,p,q,\beta}=\frac{1}{\Gamma\left(\frac{\beta}{2}p\left(n-q\right)+1\right)}\prod_{i=1}^{p}\frac{\Gamma\left(\frac{\beta}{2}\left(n-i+1\right)\right)}{\Gamma\left(\frac{\beta}{2}\left(q-i+1\right)\right)}\]
after some simplifications.

The constant $c_{n,p,q,\beta}^{\left(1\right)}$ can be calculated
in a similar way. Lemma \ref{lem:2nd-beta-integral} implies that
\begin{eqnarray*}
 &  & \underset{D_{1}}{\int\cdots\int}x_{1}\left|\Delta\left(\mathbf{x}\right)\right|^{\beta}\prod_{i=1}^{p}\left(\left(x_{i}\right)^{\frac{\beta}{2}\left(n-p-q+1\right)-1}dx_{i}\right)\\
 &  & =\frac{\frac{\beta}{2}\left(n-p-q+1\right)+\frac{\beta}{2}\left(p-1\right)}{\Gamma\left(\frac{\beta}{2}p\left(n-q\right)+2\right)}\prod_{i=1}^{p}\frac{\Gamma\left(1+\frac{\beta}{2}i\right)\Gamma\left(\frac{\beta}{2}\left(n-q-i+1\right)\right)}{\Gamma\left(1+\frac{\beta}{2}\right)}\\
 &  & =\frac{\frac{\beta}{2}\left(n-q\right)}{\frac{\beta}{2}p\left(n-q\right)+1}\left(\frac{1}{\Gamma\left(\frac{\beta}{2}p\left(n-q\right)+1\right)}\prod_{i=1}^{p}\frac{\Gamma\left(1+\frac{\beta}{2}i\right)\Gamma\left(\frac{\beta}{2}\left(n-q-i+1\right)\right)}{\Gamma\left(1+\frac{\beta}{2}\right)}\right).\end{eqnarray*}
Therefore, \[
c_{n,p,q,\beta}^{\left(1\right)}=-\left(\frac{\beta}{2}\left(q-p+1\right)-1\right)\frac{\frac{\beta}{2}p\left(n-q\right)}{\frac{\beta}{2}p\left(n-q\right)+1},\]
which completes the proof of the $1\leq p\leq q\leq\frac{n}{2}$ case.

$ $

\subsubsection{$\frac{n}{2}\leq p\leq q\leq n$ case}

This computation is closely related to that for the $1\leq p\leq q\leq\frac{n}{2}$
case.

To see the connection between the $\frac{n}{2}\leq p\leq q\leq n$
and $1\leq p\leq q\leq\frac{n}{2}$ cases, we define the generator
matrix and the orthogonal complement plane. For any given plane $P\in\mathcal{G}_{n,p}\left(\mathbb{L}\right)$,
the generator matrix $\mathbf{P}\in\mathbb{L}^{n\times p}$ is the
matrix whose $p$ columns are orthonormal and expand the plane $P$.
The generator matrix is not uniquely defined. However, the chordal
distance between $P\in\mathcal{G}_{n,p}\left(\mathbb{L}\right)$ and
$Q\in\mathcal{G}_{n,q}\left(\mathbb{L}\right)$ can be uniquely defined
by their generator matrices. Indeed, \[
d_{c}^{2}\left(P,Q\right)=\min\left(p,q\right)-\mathrm{tr}\left(\mathbf{P}^{\dagger}\mathbf{Q}\mathbf{Q}^{\dagger}\mathbf{P}\right),\]
where $\mathbf{P}$ and $\mathbf{Q}$ are generator matrices for the
plane $P$ and $Q$ respectively. It can be shown that the chordal
distance is independent of the choice of the generator matrices. The
orthogonal complement plane is defined as follows. For any given plane
$P\in\mathcal{G}_{n,p}\left(\mathbb{L}\right)$, its orthogonal complement
plane $P^{\perp}$ is the plane in $\mathcal{G}_{n,n-p}\left(\mathbb{L}\right)$
such that the minimum principle angle between $P$ and $P^{\perp}$
is $\frac{\pi}{2}$. It is straightforward that $\mathbf{P}^{\dagger}\mathbf{P}^{\perp}=\mathbf{0}$
where $\mathbf{P}$ and $\mathbf{P}^{\perp}$ are the generator matrices
for $P$ and $P^{\perp}$ respectively, and the matrix $\mathbf{0}$
is the $p\times\left(n-p\right)$ matrix with all elements $0$. 

With the definition of the orthogonal complement plane, the chordal
distance between $P$ and $Q$ can be related to that between $P^{\perp}$
and $Q^{\perp}$. The relationship is given in the following lemma. 

\begin{lemma}
\label{lem:orthogonal-complement}For any given planes $P\in\mathcal{G}_{n,p}\left(\mathbb{L}\right)$
and $Q\in\mathcal{G}_{n,q}\left(\mathbb{L}\right)$, let $P^{\perp}\in\mathcal{G}_{n,n-p}\left(\mathbb{L}\right)$
and $Q^{\perp}\in\mathcal{G}_{n,n-q}\left(\mathbb{L}\right)$ be their
orthogonal complement planes respectively. Then \[
d_{c}^{2}\left(P,Q\right)=d_{c}^{2}\left(P^{\perp},Q^{\perp}\right).\]

\end{lemma}
\begin{proof}
This lemma can be proved by the generator matrices. Let $\mathbf{P}$,
$\mathbf{Q}$, $\mathbf{P}^{\perp}$ and $\mathbf{Q}^{\perp}$ be
the generator matrices for $P$, $Q$, $P^{\perp}$ and $Q^{\perp}$
respectively. Without loss of generality, we also assume that $1\leq p\leq q\leq n$.
Then\begin{eqnarray*}
p & = & \mathrm{tr}\left(\mathbf{P}^{\dagger}\left[\mathbf{Q}\mid\mathbf{Q}^{\perp}\right]\left[\mathbf{Q}\mid\mathbf{Q}^{\perp}\right]^{\dagger}\mathbf{P}\right)\\
 & = & \mathrm{tr}\left(\mathbf{P}^{\dagger}\mathbf{Q}\mathbf{Q}^{\dagger}\mathbf{P}\right)+\mathrm{tr}\left(\mathbf{P}^{\dagger}\mathbf{Q}^{\perp}\left(\mathbf{Q}^{\perp}\right)^{\dagger}\mathbf{P}\right),\end{eqnarray*}
where the matrix $\left[\mathbf{Q}\mid\mathbf{Q}^{\perp}\right]$
is the one composed of $\mathbf{Q}$ and $\mathbf{Q}^{\perp}$. Similarly,
\begin{eqnarray*}
n-q & = & \mathrm{tr}\left(\left(\mathbf{Q}^{\perp}\right)^{\dagger}\left[\mathbf{P}\mid\mathbf{P}^{\perp}\right]\left[\mathbf{P}\mid\mathbf{P}^{\perp}\right]^{\dagger}\mathbf{Q}^{\perp}\right)\\
 & = & \mathrm{tr}\left(\left(\mathbf{Q}^{\perp}\right)^{\dagger}\mathbf{P}\mathbf{P}^{\dagger}\mathbf{Q}^{\perp}\right)+\mathrm{tr}\left(\left(\mathbf{Q}^{\perp}\right)^{\dagger}\mathbf{P}^{\perp}\left(\mathbf{P}^{\perp}\right)^{\dagger}\mathbf{Q}^{\perp}\right).\end{eqnarray*}
Then \begin{eqnarray*}
d_{c}^{2}\left(P,Q\right) & \overset{\left(a\right)}{=} & p-\mathrm{tr}\left(\mathbf{P}^{\dagger}\mathbf{Q}\mathbf{Q}^{\dagger}\mathbf{P}\right)\\
 & = & \mathrm{tr}\left(\mathbf{P}^{\dagger}\mathbf{Q}^{\perp}\left(\mathbf{Q}^{\perp}\right)^{\dagger}\mathbf{P}\right)\\
 & = & n-q-\mathrm{tr}\left(\left(\mathbf{Q}^{\perp}\right)^{\dagger}\mathbf{P}^{\perp}\left(\mathbf{P}^{\perp}\right)^{\dagger}\mathbf{Q}^{\perp}\right)\\
 & \overset{\left(b\right)}{=} & d_{c}^{2}\left(P^{\perp},Q^{\perp}\right),\end{eqnarray*}
where (a) and (b) are from the definition of the chordal distance
and the facts that $\min\left(p,q\right)=p$ and $\min\left(n-p,n-q\right)=n-q$.
\end{proof}

By this lemma, the connection between the $\frac{n}{2}\leq p\leq q\leq n$
case and $1\leq p\leq q\leq\frac{n}{2}$ case is clear. The volume
formula for the $\frac{n}{2}\leq p\leq q\leq n$ case can be calculated
as follows. \begin{eqnarray*}
\mu\left(B_{P}\left(\delta\right)\right) & = & \Pr\left(Q\in\mathcal{G}_{n,q}\left(\mathbb{L}\right):\; d_{c}^{2}\left(P,Q\right)\leq\delta^{2}\right)\\
 & = & \Pr\left(Q^{\perp}\in\mathcal{G}_{n,n-q}\left(\mathbb{L}\right):\; d_{c}^{2}\left(P^{\perp},Q^{\perp}\right)\leq\delta^{2}\right)\\
 & = & \mu\left(B_{P^{\perp}}\left(\delta\right)\right)\end{eqnarray*}
where $B_{P}\left(\delta\right)$ and $B_{P^{\perp}}\left(\delta\right)$
are the metric balls in $\mathcal{G}_{n,q}\left(\mathbb{L}\right)$
and $\mathcal{G}_{n,n-q}\left(\mathbb{L}\right)$ respectively. Therefore,
the results for the $1\leq p\leq q\leq\frac{n}{2}$ case can be directly
applied by letting $p'=n-q$ and $q'=n-p$. Finally after some simplification,
we have \begin{eqnarray*}
\mu\left(B_{P}\left(\delta\right)\right) & = & c_{n,p,q,\beta}\delta^{\beta p\left(n-q\right)}\left(1+c_{n,p,q,\beta}^{\left(1\right)}\delta^{2}+o\left(\delta^{2}\right)\right),\end{eqnarray*}
where \[
c_{n,p,q,\beta}=\frac{1}{\Gamma\left(\frac{\beta}{2}p\left(n-q\right)+1\right)}\prod_{i=1}^{n-q}\frac{\Gamma\left(\frac{\beta}{2}\left(n-i+1\right)\right)}{\Gamma\left(\frac{\beta}{2}\left(n-p-i+1\right)\right)},\]
and \[
c_{n,p,q,\beta}^{\left(1\right)}=-\left(\frac{\beta}{2}\left(q-p+1\right)-1\right)\frac{\frac{\beta}{2}p\left(n-q\right)}{\frac{\beta}{2}p\left(n-q\right)+1}.\]

$ $

\subsubsection{$1\leq p\leq\frac{n}{2}\leq q\leq n$ case}

This computation is again related to that for the $1\leq p\leq q\leq\frac{n}{2}$
case.

Similar to the $\frac{n}{2}\leq p\leq q\leq n$ case, the connection
between the $\frac{n}{2}\leq p\leq q\leq n$ case and $1\leq p\leq q\leq\frac{n}{2}$
case can be revealed by the generator matrix and the orthogonal complement
plane. Let $p'=\min\left(p,n-q\right)$ and $q'=\max\left(p,n-q\right)$.
Then $1\leq p'\leq q'\leq\frac{n}{2}$. For any given planes $P\in\mathcal{G}_{n,p}\left(\mathbb{L}\right)$
and $Q\in\mathcal{G}_{n,q}\left(\mathbb{L}\right)$, let $Q^{\perp}\in\mathcal{G}_{n,n-q}\left(\mathbb{L}\right)$
be the orthogonal complement plane of $Q$. Let $\mathbf{P}$, $\mathbf{Q}$
and $\mathbf{Q}^{\perp}$ be the generator matrices for $P$, $Q$
and $Q^{\perp}$. Then \begin{eqnarray*}
d_{c}^{2}\left(P,Q\right) & = & p-\mathrm{tr}\left(\mathbf{P}^{\dagger}\mathbf{Q}\mathbf{Q}^{\dagger}\mathbf{P}\right)\\
 & = & \mathrm{tr}\left(\mathbf{P}^{\dagger}\mathbf{Q}^{\perp}\left(\mathbf{Q}^{\perp}\right)^{\dagger}\mathbf{P}\right)\\
 & = & p'-d_{c}^{2}\left(P,Q^{\perp}\right).\end{eqnarray*}
Therefore, \begin{eqnarray*}
\mu\left(B_{P}\left(\delta\right)\right) & = & \Pr\left(Q\in\mathcal{G}_{n,q}\left(\mathbb{L}\right):\; d_{c}^{2}\left(P,Q\right)\leq\delta^{2}\right)\\
 & = & \Pr\left(Q^{\perp}\in\mathcal{G}_{n,n-q}\left(\mathbb{L}\right):\; d_{c}^{2}\left(P,Q^{\perp}\right)\geq p-\delta^{2}\right).\end{eqnarray*}

Now calculate the volume formula. Note that \[
\mu\left(B_{P}\left(\delta\right)\right)=\Pr\left(Q^{\perp}\in\mathcal{G}_{n,n-q}\left(\mathbb{L}\right):\; d_{c}^{2}\left(P,Q^{\perp}\right)\geq p'-\delta^{2}\right).\]
Then\begin{eqnarray*}
\mu\left(B_{P}\left(\delta\right)\right) & = & \underset{0\leq r_{i}^{2}\leq1,\;\sum_{i=1}^{p'}\left(1-r_{i}^{2}\right)\ge p'-\delta^{2}}{\int\cdots\int}d\mu_{\bm{\theta},p',q'}\\
 & = & \underset{0\leq r_{i}^{2}\leq1,\;\sum_{i=1}^{p'}r_{i}^{2}\leq\delta^{2}}{\int\cdots\int}d\mu_{\bm{\theta},p',q'},\end{eqnarray*}
where $d\mu_{\bm{\theta},p',q'}$ is the invariant measure with parameter
$n$, $p'$ and $q'$. Substitute the form for $d\mu_{\bm{\theta},p',q'}$
(\ref{eq:d_mu_0}) into the above formula. Then\begin{eqnarray*}
\mu\left(B_{P}\left(\delta\right)\right) & = & v_{n,p',q',\beta}\delta^{\beta p'q'}\underset{D_{1}}{\int\cdots\int}\left|\Delta\left(\mathbf{x}\right)\right|^{\beta}\prod_{i=1}^{p'}\left(x_{i}^{\frac{\beta}{2}\left(q'-p'+1\right)-1}\left(1-\delta^{2}x_{i}\right)^{\frac{\beta}{2}\left(n-p'-q'+1\right)-1}dx_{i}\right)\\
 & = & c_{n,p',q',\beta}\delta^{\beta p'q'}\left(1+c_{n,p',q',\beta}^{\left(1\right)}\delta^{2}+o\left(\delta^{2}\right)\right),\end{eqnarray*}
where the first equation comes from the variable changes $\delta^{2}x_{i}=r_{i}^{2}$
$\left(1\leq i\leq p'\right)$, $v_{n,p',q',\beta}$ is defined in
(\ref{eq:v_n_p_q_beta}),\[
D_{1}\triangleq\left\{ \mathbf{x}:\;\sum_{i=1}^{p'}x_{i}\leq1,\;0\leq x_{i}\;\mathrm{for}\; i=1,\cdots,p'\right\} ,\]
 \[
c_{n,p',q',\beta}=v_{n,p',q',\beta}\underset{D_{1}}{\int\cdots\int}\left|\Delta\left(\mathbf{x}\right)\right|^{\beta}\prod_{i=1}^{p'}\left(x_{i}^{\frac{\beta}{2}\left(q'-p'+1\right)-1}dx_{i}\right),\]
and \[
c_{n,p',q',\beta}^{\left(1\right)}=-p'\left(\frac{\beta}{2}\left(n-p'-q'+1\right)-1\right)\frac{\underset{D_{1}}{\int\cdots\int}x_{1}\left|\Delta\left(\mathbf{x}\right)\right|^{\beta}\prod_{i=1}^{p'}\left(x_{i}^{\frac{\beta}{2}\left(q'-p'+1\right)-1}dx_{i}\right)}{\underset{D_{1}}{\int\cdots\int}\left|\Delta\left(\mathbf{x}\right)\right|^{\beta}\prod_{i=1}^{p'}\left(x_{i}^{\frac{\beta}{2}\left(q'-p'+1\right)-1}dx_{i}\right)}.\]
Applying Lemma \ref{lem:2nd-beta-integral} and after some simplification,
we have that\[
c_{n,p',q',\beta}=\frac{1}{\Gamma\left(\frac{\beta}{2}p'q'+1\right)}\prod_{i=1}^{p'}\frac{\Gamma\left(\frac{\beta}{2}\left(n-i+1\right)\right)}{\Gamma\left(\frac{\beta}{2}\left(n-p'-i+1\right)\right)},\]
and \[
c_{n,p',q',\beta}^{\left(1\right)}=-\left(\frac{\beta}{2}\left(n-p'-q'+1\right)-1\right)\frac{\frac{\beta}{2}p'q'}{\frac{\beta}{2}p'q'+1}.\]

Summarily, if $1\leq p\leq\frac{n}{2}\leq q\leq n$,\[
\mu\left(B\left(\delta\right)\right)=c_{n,p,q,\beta}\delta^{\beta p\left(n-q\right)}\left(1+c_{n,p,q,\beta}^{\left(1\right)}\delta^{2}+o\left(\delta^{2}\right)\right),\]
where \[
c_{n,p,q,\beta}=\left\{ \begin{array}{ll}
\frac{1}{\Gamma\left(\frac{\beta}{2}p\left(n-q\right)+1\right)}\prod_{i=1}^{n-q}\frac{\Gamma\left(\frac{\beta}{2}\left(n-i+1\right)\right)}{\Gamma\left(\frac{\beta}{2}\left(q-i+1\right)\right)} & \mathrm{if}\; p+q\geq n\\
\frac{1}{\Gamma\left(\frac{\beta}{2}p\left(n-q\right)+1\right)}\prod_{i=1}^{p}\frac{\Gamma\left(\frac{\beta}{2}\left(n-i+1\right)\right)}{\Gamma\left(\frac{\beta}{2}\left(n-p-i+1\right)\right)} & \mathrm{if}\; p+q\leq n\end{array}\right.,\]
and \[
c_{n,p,q,\beta}^{\left(1\right)}=-\left(\frac{\beta}{2}\left(q-p+1\right)-1\right)\frac{\frac{\beta}{2}p\left(n-q\right)}{\frac{\beta}{2}p\left(n-q\right)+1}.\]

\subsection{\label{sub:Proof-of-DCF-lb}Proof of the lower bound on $D^{*}\left(K\right)$}

Assume a source $Q$ is uniformly distributed in $\mathcal{G}_{n,p}\left(\mathbb{L}\right)$.
For any codebook $\mathcal{C}$, define the empirical cumulative distribution
function as\[
F_{d_{c}^{2},\mathcal{C}}\left(x\right)=\Pr\left\{ Q:\;\left(\underset{P\in\mathcal{C}}{\min}\; d_{c}^{2}\left(P,Q\right)\right)\leq x\right\} .\]
Then the distortion associated with the codebook $\mathcal{C}$ is
given by \begin{equation}
D\left(\mathcal{C}\right)=\int_{0}^{p}x\cdot dF_{d_{c}^{2},\mathcal{C}}\left(x\right).\label{eq:distortion_for_a_codebook}\end{equation}

The following theorem gives the empirical distribution to minimize
the distortion.

\begin{lemma}
\label{lem:optimal-empirical-distribution}The empirical distribution
function minimizing the distortion for a given $K$ is \[
F_{d_{c}^{2},\mathcal{C}}^{*}\left(x\right)=\left\{ \begin{array}{ll}
0 & \mathrm{if}\; x<0\\
K\cdot\mu\left(B\left(\sqrt{x}\right)\right) & \mathrm{if}\;0\leq x\leq x^{*}\\
1 & \mathrm{if}\; x^{*}<x\end{array}\right.,\]
where $x^{*}$ satisfies $K\cdot\mu\left(B\left(\sqrt{x^{*}}\right)\right)=1$. 
\end{lemma}
\begin{proof}
For any empirical distribution $F_{d_{c}^{2},\mathcal{C}}\left(x\right)$,
\begin{eqnarray*}
F_{d_{c}^{2},\mathcal{C}}\left(x\right) & = & \Pr\left\{ Q:\;\left(\underset{P\in\mathcal{C}}{\min}\; d_{c}^{2}\left(P,Q\right)\right)\leq x\right\} \\
 & = & \Pr\left(\cup_{i=1}^{K}\left\{ Q:\; d_{c}^{2}\left(P_{i},Q\right)\leq x\right\} \right)\\
 & \leq & \sum_{i=1}^{K}\Pr\left\{ Q:\; d_{c}^{2}\left(P_{i},Q\right)\leq x\right\} \\
 & = & K\cdot\mu\left(B\left(\sqrt{x}\right)\right).\end{eqnarray*}
Thus \begin{equation}
F_{d_{c}^{2},\mathcal{C}}\left(x\right)\leq\min\left(1,K\cdot\mu\left(B\left(\sqrt{x}\right)\right)\right).\label{eq:empirical-distribution}\end{equation}
Therefore, \begin{eqnarray*}
 &  & \int_{0}^{p}x\cdot dF_{d_{c}^{2},\mathcal{C}}\left(x\right)-\int_{0}^{p}x\cdot dF_{d_{c}^{2},\mathcal{C}}^{*}\left(x\right)\\
 &  & \overset{\left(a\right)}{=}\int_{0}^{p}F_{d_{c}^{2},\mathcal{C}}^{*}\left(x\right)dx-\int_{0}^{p}F_{d_{c}^{2},\mathcal{C}}\left(x\right)dx\\
 &  & =\int_{0}^{x_{0}}\left(K\cdot\mu\left(B\left(\sqrt{x}\right)\right)-F_{d_{c}^{2},\mathcal{C}}\left(x\right)\right)dx+\int_{x_{0}}^{p}\left(1-F_{d_{c}^{2},\mathcal{C}}\left(x\right)\right)dx\\
 &  & \overset{\left(b\right)}{\geq}0,\end{eqnarray*}
where (a) follows from integration by parts, and (b) follows from
(\ref{eq:empirical-distribution}).
\end{proof}

From Lemma \ref{lem:optimal-empirical-distribution}, it is clear
that \begin{align}
D^{*}\left(K\right) & \geq\int_{0}^{p}x\cdot dF_{d_{c}^{2},\mathcal{C}}^{*}\left(x\right)\nonumber \\
 & =\int_{0}^{p}x\cdot dKF\left(x\right),\label{eq:DRF-LB-integral}\end{align}
where $F\left(x\right)\triangleq\mu\left(B\left(\sqrt{x}\right)\right)$. 

$ $

\subsubsection{Proof of Theorem \ref{thm:DRF_bounds}}

Theorem \ref{thm:DRF_bounds} is proved by substituting the volume
formula (\ref{eq:simplified_volume_formula}) into (\ref{eq:DRF-LB-integral}).
Another way to prove it is to apply the lower bound in Corollary \ref{thm:DRF_bounds_details},
whose proof is more involved and given in the following.

$ $

\subsubsection{Proof of the lower bounds in Corollary \ref{thm:DRF_bounds_details}}

The difficulty to calculate (\ref{eq:DRF-LB-integral}) is that we
don't know the exact $F\left(x\right)$ for some cases. To overcome
this difficulty, we construct a further lower bound on (\ref{eq:DRF-LB-integral}). 

For all cases except the $\beta=1$ and $q=p$ case, a lower bound
on (\ref{eq:DRF-LB-integral}) is constructed as follows. Let $F_{0}\left(x\right)=c_{n,p,q,\beta}x^{\frac{\beta}{2}p\left(n-q\right)}$
and $x_{0}$ satisfy $KF_{0}\left(x_{0}\right)=1$. Since $F\left(x\right)\leq F_{0}\left(x\right)$
(Corollary \ref{cor:volume-bounds}), $KF\left(x_{0}\right)\leq KF_{0}\left(x_{0}\right)=1$.
But $KF\left(x^{*}\right)=1$. We have $x_{0}\leq x^{*}$. Therefore,
\begin{align*}
 & \int_{0}^{x^{*}}x\cdot dKF\left(x\right)\\
 & =\int_{0}^{x^{*}}\left(1-KF\left(x\right)\right)dx\\
 & \geq\int_{0}^{x_{0}}\left(1-KF\left(x\right)\right)dx\\
 & \geq\int_{0}^{x_{0}}\left(1-KF_{0}\left(x\right)\right)dx\\
 & =\int_{0}^{x_{0}}x\cdot dKF_{0}\left(x\right)\\
 & =\frac{\beta p\left(n-q\right)}{\beta p\left(n-q\right)+2}\left(c_{n,p,q,\beta}K\right)^{-\frac{2}{\beta p\left(n-q\right)}}.\end{align*}

For the case $\beta=1$ and $q=p$, the computation is more complicated.
The following lemma is helpful. 

\begin{lemma}
\label{lem:DRF-LB-LB}Let $\alpha=\frac{1}{2}p\left(n-p\right)$.
Let $F_{0}\left(x\right)=c_{n,p,p,1}x^{\alpha}$ and $x_{0}$ satisfy
$KF_{0}\left(x_{0}\right)=1$. Let $F_{ub}=c_{n,p,p,1}x^{\alpha}\left(1-x\right)^{-\frac{p}{2}}$
and $x_{ub}$ satisfy $KF_{ub}\left(x_{ub}\right)=1$. Let $F_{ubub}=c_{n,p,p,1}x^{\alpha}\left(1-x_{0}\right)^{-\frac{p}{2}}$
and $x_{ubub}$ satisfy $KF_{ubub}\left(x_{ubub}\right)=1$. Then
\begin{align*}
\int_{0}^{x_{ubub}}x\cdot dKF_{ubub}\left(x\right) & \leq\int_{0}^{x_{ub}}x\cdot dKF_{ub}\left(x\right)\leq\int_{0}^{x^{*}}x\cdot dKF\left(x\right).\end{align*}

\end{lemma}
\begin{proof}
Similar to the arguments for all the cases except the $\beta=1$ and
$q=p$ case, it can be proved that $x_{ub}\leq x^{*}\leq x_{0}$ and
$\int_{0}^{x_{ub}}x\cdot dKF_{ub}\left(x\right)\leq\int_{0}^{x^{*}}x\cdot dKF\left(x\right)$.
Then $\left(1-x\right)^{-\frac{p}{2}}\leq\left(1-x_{0}\right)^{-\frac{p}{2}}$
for $x\in\left[0,x_{ub}\right]$. It implies $F_{ub}\left(x\right)\leq F_{ubub}\left(x\right)$
for $x\in\left[0,x_{ub}\right]$. Therefore, $x_{ubub}\leq x_{ub}$
and $\int_{0}^{x_{ubub}}x\cdot dKF_{ubub}\left(x\right)\leq\int_{0}^{x_{ub}}x\cdot dKF_{ub}\left(x\right)$. 
\end{proof}

We calculate $\int_{0}^{x_{ubub}}x\cdot dKF_{ubub}\left(x\right)$
as follows. $x_{ubub}=x_{0}\left(1-x_{0}\right)^{\frac{1}{n-p}}$.
\begin{align*}
 & \int_{0}^{x_{ubub}}x\cdot dKF_{ubub}\left(x\right)\\
 & =\int_{0}^{x_{ubub}}x\cdot dKF_{0}\left(x\right)\left(1-x_{0}\right)^{-\frac{p}{2}}\\
 & =\frac{p\left(n-p\right)}{p\left(n-p\right)+2}Kc_{n,p,p,1}\left(1-x_{0}\right)^{-\frac{p}{2}}x_{ubub}^{\frac{1}{2}p\left(n-p\right)+1}\\
 & =\frac{p\left(n-p\right)}{p\left(n-p\right)+2}x_{ubub}\\
 & =\frac{p\left(n-p\right)}{p\left(n-p\right)+2}\left(c_{n,p,p,1}K\right)^{-\frac{2}{p\left(n-p\right)}}\left(1-\left(c_{n,p,p,1}K\right)^{-\frac{2}{p\left(n-p\right)}}\right)^{\frac{1}{n-p}}.\end{align*}

\subsection{\label{sub:Proof-of-DCF-ub}Proof of the upper bound on $D^{*}\left(K\right)$}

To get an upper bound on $D^{*}\left(K\right)$, we shall compute
the average distortion of the random codes. Let $\mathcal{C}_{\mathrm{rand}}=\left\{ P_{1},\cdots,P_{K}\right\} $
be a random code whose codewords $P_{i}$'s are independently drawn
from the uniform distribution on $\mathcal{G}_{n,p}\left(\mathbb{L}\right)$.
For any given element $Q\in\mathcal{G}_{n,p}\left(\mathbb{L}\right)$,
define $X_{i}=d_{c}^{2}\left(P_{i},Q\right)$, $1\leq i\leq K$. Then
$X_{i}$'s are independent and identically distributed (i.i.d.) random
variables with distribution function \[
F\left(x\right)=\mu\left(B\left(\sqrt{x}\right)\right)=c_{n,p,p,\beta}x^{\frac{t}{2}}\left(1+O\left(x\right)\right).\]
 Define $W_{K}=\min\left(X_{1},\cdots,X_{K}\right)$. Then \begin{eqnarray*}
 &  & \mathrm{E}_{\mathcal{C}_{\mathrm{rand}}}\left[D\left(\mathcal{C}_{\mathrm{rand}}\right)\right]\\
 &  & =\mathrm{E}_{\mathcal{C}_{\mathrm{rand}}}\left[\mathrm{E}_{Q}\left[\underset{P_{i}\in\mathcal{C}_{\mathrm{rand}}}{\min}\; d_{c}^{2}\left(Q,P_{i}\right)\right]\right]\\
 &  & =\mathrm{E}_{Q}\left[\mathrm{E}_{\mathcal{C}_{\mathrm{rand}}}\left[\underset{P_{i}\in\mathcal{C}_{\mathrm{rand}}}{\min}\; d_{c}^{2}\left(Q,P_{i}\right)\right]\right]\\
 &  & =\mathrm{E}_{Q}\left[\mathrm{E}_{W_{K}}\left[W_{K}\right]\right].\end{eqnarray*}

To calculate $\mathrm{E}_{W_{K}}\left[W_{K}\right]$, we need to know
the distribution of $W_{K}$. To derive it, the the following lemma
is useful.

\begin{lemma}
\label{lem:W_lb_ub}Let $X_{i}$'s $1\leq i\leq K$ be i.i.d. random
variables with distribution function $F\left(x\right)$. Let $W_{K}=\min\left(X_{1},\cdots,X_{K}\right)$.
Then \begin{eqnarray*}
\exp\left(-KF\left(x\right)\right) & > & \Pr\left(W_{K}>x\right)=\left(1-F\left(x\right)\right)^{K}\end{eqnarray*}
where the upper bound holds for all $x$.
\end{lemma}
\begin{proof}
See \cite[page 10]{JanosGalambos1987_extreme_order_statistics}. 
\end{proof}

With the above upper bound on the distribution function of $W_{K}$,
we derive an upper bound on $\mathrm{E}_{W_{K}}\left[W_{K}\right]$.
In the following, we use $\mathrm{E}\left[\cdot\right]$ instead of
$\mathrm{E}_{W_{K}}\left[\cdot\right]$ for simplicity. Let $F_{lb}\left(x\right)$
be an arbitrary distribution function such that $F_{lb}\left(x\right)\leq F\left(x\right)$.
It is clear that $F_{lb}\left(x\right)$ is zero if $x<0$. Then\begin{align*}
\mathrm{E}\left[W_{K}\right] & =\int_{0}^{p}\Pr\left(W_{K}>x\right)dx\\
 & \overset{\left(a\right)}{\leq}\int_{0}^{p}\exp\left(-KF\left(x\right)\right)dx\\
 & \leq\int_{0}^{p}\exp\left(-KF_{lb}\left(x\right)\right)dx\\
 & \leq\int_{0}^{x_{0}}\exp\left(-KF_{lb}\left(x\right)\right)dx+\int_{x_{0}}^{p}\exp\left(-KF_{lb}\left(x_{0}\right)\right)dx\\
 & \leq\int_{0}^{x_{0}}\exp\left(-KF_{lb}\left(x\right)\right)dx+p\exp\left(-KF_{lb}\left(x_{0}\right)\right),\end{align*}
where (a) follows Lemma \ref{lem:W_lb_ub}. Here and throughout, $x_{0}=\left(c_{n,p,q,\beta}K\right)^{-\frac{2}{\beta p\left(n-q\right)}a}$,
$a\in\left(0,1\right)$ is an arbitrary real number and $K$ is large
enough to guarantee $x_{0}\leq1$.

If 1) $\mathbb{L}=\mathbb{R}$ and $p\leq q\leq p+1$, or 2) $\mathbb{L}=\mathbb{C}$
and $q=p$, then we can take $F_{lb}\left(x\right)=c_{n,p,q,\beta}x^{\frac{\beta}{2}p\left(n-q\right)}$
(Corollary \ref{cor:Exact-Volume-Formula} and \ref{cor:volume-bounds}).
We have\begin{align*}
\mathrm{E}\left[W_{K}\right] & \leq\int_{0}^{\infty}\exp\left(-Kc_{n,p,q,\beta}x^{\frac{2}{\beta p\left(n-q\right)}}\right)dx\\
 & \quad\quad+p\exp\left(-Kc_{n,p,q,\beta}\left(\left(Kc_{n,p,q,\beta}\right)^{-\frac{2}{\beta p\left(n-q\right)}a}\right)^{\frac{\beta p\left(n-q\right)}{2}}\right)\\
 & \overset{\left(b\right)}{=}\left(Kc_{n,p,q,\beta}\right)^{-\frac{2}{\beta p\left(n-q\right)}}\int_{0}^{\infty}e^{-y}\cdot dy^{\frac{1}{\alpha}}\\
 & \quad\quad+p\exp\left(-\left(Kc_{n,p,q,\beta}\right)^{1-a}\right)\\
 & =\left(Kc_{n,p,q,\beta}\right)^{-\frac{2}{\beta p\left(n-q\right)}}\frac{2\Gamma\left(\frac{2}{\beta p\left(n-q\right)}\right)}{\beta p\left(n-q\right)}+p\exp\left(-\left(Kc_{n,p,q,\beta}\right)^{1-a}\right),\end{align*}
where (b) is from the variable change $y=Kc_{n,p,q,\beta}x^{\frac{2}{\beta p\left(n-q\right)}}$.
Thus, for any given $n$, $p$ and $q$, \[
\underset{K\rightarrow\infty}{\lim}\mathrm{E}\left[K^{\frac{2}{\beta p\left(n-q\right)}}W_{K}\right]\leq\frac{2\Gamma\left(\frac{2}{\beta p\left(n-q\right)}\right)}{\beta p\left(n-q\right)}\left(c_{n,p,q,\beta}\right)^{-\frac{2}{\beta p\left(n-q\right)}},\]
and so, when $K$ is large enough, \[
D^{*}\left(K\right)\leq\frac{2\Gamma\left(\frac{2}{\beta p\left(n-q\right)}\right)}{\beta p\left(n-q\right)}\left(c_{n,p,q,\beta}K\right)^{-\frac{2}{\beta p\left(n-q\right)}}\left(1+o\left(1\right)\right).\]

If 1) $\mathbb{L}=\mathbb{R}$ and $q\geq p+1$, or 2) $\mathbb{L}=\mathbb{C}$
and $q>p$, then $F\left(x\right)\geq c_{n,p,q,\beta}x^{\alpha}\left(1-x\right)^{\gamma}$
where $\alpha=\frac{\beta}{2}p\left(n-q\right)$ and $\gamma=\frac{\beta p\left(q-p+1\right)}{2}-p$
(Corollary \ref{cor:Exact-Volume-Formula} and \ref{cor:volume-bounds}).
Note that $\left(1-x\right)^{\gamma}\geq\left(1-x_{0}\right)^{\gamma}$
for all $x\in\left[0,x_{0}\right]$. We take $F_{lb}\left(x\right)=c_{n,p,q,\beta}x^{\alpha}\left(1-x_{0}\right)^{\gamma}$.
Then\begin{align*}
\mathrm{E}\left[W_{K}\right] & \leq\int_{0}^{x_{0}}\exp\left(-Kc_{n,p,q,\beta}x^{\alpha}\left(1-x_{0}\right)^{\gamma}\right)dx\\
 & \quad\quad+p\exp\left(-Kc_{n,p,q,\beta}x_{0}^{\alpha}\left(1-x_{0}\right)^{\gamma}\right)dx\\
 & \overset{\left(c\right)}{\leq}\left(Kc_{n,p,q,\beta}\right)^{-\frac{1}{\alpha}}\left(1-\left(Kc_{n,p,q,\beta}\right)^{-\frac{a}{\alpha}}\right)^{-\frac{\gamma}{\alpha}}\int_{0}^{\infty}e^{-y}\cdot dy^{\frac{1}{\alpha}}\\
 & \quad\quad p\exp\left(-\left(Kc_{n,p,q,\beta}\right)^{1-a}\left(1-\left(Kc_{n,p,q,\beta}\right)^{-\frac{a}{\alpha}}\right)^{\gamma}\right)\\
 & =\frac{\Gamma\left(\frac{1}{\alpha}\right)}{\alpha}\left(Kc_{n,p,q,\beta}\right)^{-\frac{1}{\alpha}}\left(1-\left(Kc_{n,p,q,\beta}\right)^{-\frac{a}{\alpha}}\right)^{-\frac{\gamma}{\alpha}}\\
 & \quad\quad+p\exp\left(-\left(Kc_{n,p,q,\beta}\right)^{1-a}\left(1-\left(Kc_{n,p,q,\beta}\right)^{-\frac{a}{\alpha}}\right)^{\gamma}\right),\end{align*}
where (c) follows the variable change $y=Kc_{n,p,q,\beta}x^{\alpha}\left(1-x_{0}\right)^{\gamma}$.
Once more, for any given $n$, $p$ and $q$, \[
\underset{K\rightarrow\infty}{\lim}\mathrm{E}\left[K^{\frac{2}{\beta p\left(n-q\right)}}W_{K}\right]\leq\frac{2\Gamma\left(\frac{2}{\beta p\left(n-q\right)}\right)}{\beta p\left(n-q\right)}\left(c_{n,p,q,\beta}\right)^{-\frac{2}{\beta p\left(n-q\right)}},\]
and \[
D^{*}\left(K\right)\leq\frac{2\Gamma\left(\frac{2}{\beta p\left(n-q\right)}\right)}{\beta p\left(n-q\right)}\left(c_{n,p,q,\beta}K\right)^{-\frac{2}{\beta p\left(n-q\right)}}\left(1+o\left(1\right)\right),\]
for sufficiently large $K$. 

\newpage
\bibliographystyle{IEEEtran}
\bibliography{Bib/_Heath,Bib/_Liu_Dai,Bib/_love,Bib/_Rao,Bib/_Blum,Bib/FeedbackMIMO_append,Bib/MIMO_basic,Bib/RandomMatrix}

\newpage

\newpage

\newpage

\end{document}